\documentclass[manuscript]{acmart}

\AtBeginDocument{%
  }

\setcopyright{acmlicensed}
\copyrightyear{2018}
\acmYear{2018}
\acmDOI{XXXXXXX.XXXXXXX}

\acmJournal{TOIS}
\acmVolume{1}
\acmNumber{1}
\acmArticle{1}
\acmMonth{5}

\acmISBN{978-1-4503-XXXX-X/18/06}

\usepackage{algorithm}
\usepackage{algorithmic}
\usepackage{amsmath}
\usepackage{graphicx}
\usepackage{subcaption}
\usepackage{color}
\usepackage{multirow}
\usepackage{pgfplots}
\usepackage{svg}
\usepackage{caption}
\usepackage{subcaption}
\usepackage{graphicx}
\usepackage[normalem]{ulem}
\usepackage{soul}

\begin{document}

\title{Meta Learning to Rank for Sparsely Supervised Queries}

\author{Xuyang Wu}
\authornote{This work was completed as part of a summer internship program at Walmart Global Tech.}
\affiliation{%
  \institution{Santa Clara University}
  \streetaddress{500 El Camino Real, Santa Clara, CA 95053}
  \city{Santa Clara}
  \state{CA}
  \country{USA}
  \postcode{95053}
}
\orcid{0000-0002-8807-0016}
\email{xwu5@scu.edu}

\author{Ajit Puthenputhussery}
\email{ajit.puthenputhussery@walmart.com}
\orcid{0000-0001-7141-1534}
\affiliation{%
  \institution{Walmart Global Tech}
  \streetaddress{860 W California Ave, Sunnyvale, CA 94086}
  \city{Sunnyvale}
  \state{CA}
  \country{USA}
  \postcode{94086}
}

\author{Hongwei Shang}
\orcid{0009-0005-9856-9178}
\affiliation{%
  \institution{Walmart Global Tech}
  \streetaddress{860 W California Ave, Sunnyvale, CA 94086}
  \city{Sunnyvale}
  \state{CA}
  \country{USA}
  \postcode{94086}}
\email{hongwei.shang@walmart.com}

\author{Changsung Kang}
\orcid{0009-0007-5305-8256}
\affiliation{%
  \institution{Walmart Global Tech}
  \streetaddress{860 W California Ave, Sunnyvale, CA 94086}
  \city{Sunnyvale}
  \state{CA}
  \country{USA}
  \postcode{94086}}
 \email{changsung.kang@walmart.com} 

\author{Yi Fang}
\orcid{0000-0001-6572-4315}
\authornote{Corresponding author.}
\affiliation{%
  \institution{Santa Clara University}
  \streetaddress{500 El Camino Real, Santa Clara, CA 95053}
  \city{Santa Clara}
  \state{CA}
  \country{USA}
  \postcode{95053}}
\email{yfang@scu.edu}

\renewcommand{\shortauthors}{X. Wu, et al.}

\begin{abstract}
Supervisory signals are a critical resource for training learning to rank models. In many real-world search and retrieval scenarios, these signals may not be readily available or could be costly to obtain for some queries. The examples include domains where labeling requires professional expertise, applications with strong privacy constraints, and user engagement information that are too scarce. We refer to these scenarios as sparsely supervised queries which pose significant challenges to traditional learning to rank models.
In this work, we address sparsely supervised queries by proposing a novel meta learning to rank framework which leverages fast learning and adaption capability of meta-learning. The proposed approach accounts for the fact that different queries have different optimal parameters for their rankers, in contrast to traditional learning to rank models which only learn a global ranking model applied to all the queries. In consequence, the proposed method would yield significant advantages especially when new queries are of different characteristics with the training queries. Moreover, the proposed meta learning to rank framework is generic and flexible. We conduct a set of comprehensive experiments on both public datasets and a real-world e-commerce dataset. The results demonstrate that the proposed meta-learning approach can significantly enhance the performance of learning to rank models with sparsely labeled queries.
\end{abstract}





\maketitle

\section{Introduction}
Learning to rank (LTR), which refers to machine learning techniques on automatically constructing a model from data for ranking in search, has been widely used in modern search engines \cite{liu2009learning}. Typically, LTR involves creating a single ranking function that applies universally to all queries to order items based on their relevance. The global ranking model is generally efficient and scalable since it can be reused without requiring separate training or tuning for each query. Such an approach often delivers robust average performance and is easier to maintain in practice, making it widely adopted in LTR. However, the global ranking approach may be suboptimal for individual queries as it tends to overlook query specificity and user intent. This is particularly problematic given that relevant documents for different queries can have varying distributions in the feature space, which a global ranking function might not adequately capture \cite{DBLP:conf/sigir/AiBGC18}. For instance, considering two ranking features such as word matching and freshness, in queries like ``running shoes for flat feet'', emphasis may be placed on word matching over freshness, whereas queries like ``latest video games'' would prioritize freshness. This variation necessitates the development of query-specific rankers, as global models may not able to generalize across diverse queries. Different queries prioritize different features, leading to domain shifts that can undermine the effectiveness of models trained on different types of data. Query-specific models are desired in certain search scenarios where the characteristics of queries and user intents may lead to distinct distributions of relevant documents in the feature space, and offer the advantage of tailoring model parameters to optimize retrieval for individual queries. The prior works in the literature \cite{DBLP:conf/sigir/AiBGC18, DBLP:conf/sigir/CanCM14, DBLP:conf/sigir/GengLQALS08} have also advocated for constructing ranking functions on a per-query basis, recognizing the limitations of a global ranking function.

Moreover, learning an effective ranking function often relies on the availability of a large amount of labeled examples. It may be difficult to obtain sufficient labeled examples for many queries in the real world such as domains where labeling requires professional expertise (e.g., biomedical and legal search) and applications with strong privacy constraints \cite{DBLP:conf/sigir/WangBMN16} (e.g., personal and enterprise search). User engagement data such as clicks/add-to-cart/purchase on e-commerce platforms is indicative of individual users' relevance judgments and is relatively easy to collect with low cost, but queries with sparse user interaction are still frequently encountered on these platforms such as queries for new products and tail queries. Additionally, due to certain biases in data collection and the limited availability of labeled data, user interactions labels may not necessarily align with actual user preferences \cite{DBLP:journals/ftir/Sanderson10}. We refer to the above scenarios where queries have limited supervisory signals for learning to rank as sparsely supervised queries. 

Sparsely supervised queries pose significant challenges to LTR models, especially when learning query-specific rankers. First of all, traditional LTR methods typically require a large amount of supervised data to optimize different ranking objectives, but this design is not intended to learn ``fast'' from limited data. Although some recent works \cite{DBLP:conf/icml/YueJ09, DBLP:conf/wsdm/HofmannSWR13} have attempted to dynamically adjust the ranker's optimization direction using online LTR with historical data and current real-time data, these approaches often suffer from insufficient optimization efficiency, unmeasurable performance, or performance that is inferior to offline approaches \cite{DBLP:conf/ecir/OosterhuisSR16}. Moreover, even if an LTR model is trained with a large amount of supervisory signals, when it encounters sparsely supervised queries at run time, it may not be able to generalize well. The scarcity or limited number of examples can have a significant impact on inductive bias \cite{DBLP:journals/jair/Baxter00}. The characteristics of sparsely supervised queries could be quite different from those of the training queries, which may lead to the domain shift problem from training to prediction/inference. In addition, sparsely supervised queries usually result in a high imbalance between positive labels and negative labels since irrelevant documents can often be sampled from the dataset while relevant documents have to be labeled. There exist some works in the literature that attempted to address the above respective challenges by generating synthetic data or duplicating existing data to provide more informative training sets. For example, data augmentation \cite{DBLP:conf/aaai/YuL19, DBLP:conf/cikm/QiuJCZ21}, resampling methods \cite{DBLP:conf/ijcnn/BatuwitaP10, DBLP:journals/jair/ChawlaBHK02} and ensemble methods \cite{DBLP:conf/mcs/Dietterich00} were utilized to alleviate data sparsity, balance relevance labels and attempt to learn an unbiased model in training. These methods have limited improvement in terms of model generalization due to insufficient data and domain shift issues. And the small sample size and uneven distribution of labels can result in bias or difficulties in transferring knowledge, as well as slow adaptation to new queries, ultimately limiting generalization. On the other hand, some works aim to mitigate the impact of noise and bias through unbiased modeling perspectives and model adjustments \cite{DBLP:journals/ijdsa/YuanGSLJ23, DBLP:conf/wsdm/JoachimsSS17, DBLP:conf/wsdm/CraswellZTR08, DBLP:conf/wsdm/WangGBMN18, DBLP:conf/sigir/JoachimsGPHG05, DBLP:conf/www/AgarwalWLBN19, DBLP:conf/sigir/OosterhuisR20, DBLP:journals/tois/Oosterhuis23}. Generally, these methods model position bias by requiring extensive click logs. For instance, to optimize position bias in the data, the concept of counterfactual Inverse Propensity Scoring (IPS) was introduced in \cite{DBLP:conf/www/AgarwalWLBN19}. In our research scenario, there is a lack of positive sample data, which increases the difficulty of modeling bias. Additionally, due to insufficient training samples, minor propensities, and a large number of noisy clicks, counterfactual LTR systems frequently suffer from excessive variance. \citet{DBLP:journals/tois/Oosterhuis23} proposed the DR estimator, which provides enormous decreases in variance. These models have achieved remarkable success in the unbiased LTR field by using more efficient estimators to correct the bias problem. Last but not least, the presence of sparsely supervised queries complicates the development of query-specific rankers, as the straightforward approach of training individual models for each query would only exacerbate data sparsity and render the process infeasible.

Given these challenges, we turn to meta-learning \cite{DBLP:conf/icml/FinnAL17}, which has demonstrated its great success in the setting of few-shot learning where a model can quickly adapt to a new task using only a few data points and training iterations, as shown in a wide range of machine learning applications including image classification, dialog generation \cite{DBLP:conf/acl/QianY19}, classification \cite{DBLP:journals/corr/abs-1806-00852}, and recommendation systems \cite{DBLP:conf/kdd/LeeIJCC19, DBLP:journals/tois/CuiS0Y022, DBLP:journals/tois/HuangSYX22}. Learning to rank for sparsely supervised queries shares similar characteristics with meta-learning in a few-shot setting, because it focuses on ranking items for a query which only has a small number of labeled documents or supervisory signals. Inspired by the capabilities of meta-learning in fast learning and improving model generalization, we propose a novel meta-learning to rank approach to address sparsely supervised queries. In scenarios where labeled data is scarce and the distribution of labels is imbalanced, meta-learning can effectively utilize its efficient learning and adaptability capabilities. Moreover, meta-learning can mitigate the impact of domain shift by allowing models to quickly adapt to different data distributions through task-specific training during the learning process and fine-tuning during inference. 

In this paper, we utilize the optimization-based meta-learning approach \cite{DBLP:conf/icml/FinnAL17}, to rapidly estimate document relevance for a new query based on only a small number of labeled documents. For each query in the meta-training process, there are two sets: training set and test set. The proposed meta learning to rank model performs local and global updates. During the local update, the algorithm adjusts the parameter of the query-specific ranker on each training set (learning process). During the global update, the algorithm trains the parameter of the meta ranker to minimize the meta loss with the adapted parameters on the test sets (learning-to-learn process). Each query-specific ranker only requires few labeled instances for fine-tuning as the meta ranker is trained across all the queries and the global ranking knowledge is transferred to each query-specific ranker as initial model parameters before fine-tuning. The proposed meta-learning approach is an efficient way to learn from limited data. To estimate document relevance for a new query, the ranker can then be fine-tuned based on the limited amount of labeled documents. Due to the learning-to-learn process, the model is able to quickly adapt to a new query. Query-specific rankers enable the model to capture and adapt to the unique characteristics of each query, while the meta (global) ranker preserves scalability and efficiency across diverse queries. By leveraging the strengths of both approaches, our method aims to balance scalability with specificity, ensuring robust performance and leading to more precise and relevant results. In consequence, the proposed method leverages the fast learning and adaptation capabilities inherent in the meta-learning framework, yielding significant advantages especially when new queries are of different characteristics with the training queries. 

Long-tailed queries can be naturally tackled by the proposed meta-learning approach. A portion of these queries may only appear once, while others could appear multiple times, albeit less than a few. Our proposed meta-learning approach can handle both scenarios through without fine-tuning or with fine-tuning. For queries that appear only once, we may not use data from the same query in training. For queries that appear more frequently, we can apply fine-tuning on unseen queries. The experiments demonstrate performance improvements of the MLTR models over the baselines under both scenarios. It is worth noting that the proposed approach is not limited to long-tailed queries. Even with more frequently occurring queries, such as torso and head queries, the available labels or user engagement data could be quite scarce, especially within a short timeframe since their first appearance. To quickly learn good ranking functions for these queries is crucial for engaging users in real-world search applications. Our work is centered on fast and efficient learning from sparse labels, a focus we believe holds broad applicability across various search scenarios. The main contributions of the paper can be summarized as follow:

\begin{itemize}

    \item We propose a novel meta-learning framework for search and ranking with sparsely supervised queries. To the best of our knowledge, there is no prior work on adopting the optimization-based meta-learning for learning to rank.

    \item The proposed meta-learning to rank model can leverage its strong generalization ability during training, enabling it to sustain consistently stable performance in ranking tasks involving unseen queries.

    \item The proposed meta learning to rank model can quickly adapt to a new query with limited supervisory signals and can yield query-specific rankers with optimal model parameters for individual queries.
    
    \item The proposed approach is generic and flexible and can be applied to any existing LTR models to improve model generalization.

    \item We conduct a comprehensive set of experiments on four public learning to rank benchmarks and one real-world product search dataset. The results demonstrate the effectiveness of the proposed approach over the competitive baselines. 

\end{itemize}

\section{Related Work}

\subsection{Learning to Rank}

Learning to Rank has gained much attention from the information retrieval research and industry community, which aims to optimize the perfect search results. A series of LTR models have been developed in the research community, mainly in two divisions. One division is based on Gradient Boosted Decision Trees (GBDT), which have been used by many production search systems \citep{Yin:Hu:Tang:Rank:2016}; the other division is based on neural networks. With the intriguing interest in neural LTR models, a lot of papers have been published, and researchers keep proposing state-of-the-art methods \citep{b12,DBLP:conf/sigir/DehghaniZSKC17,Xiao:Ji:Cui:Weak:2019,Qin:Yan:Are:ICLR:2021,b12,b13,b14}. The DSSM model, cited as \cite{b12}, belongs to the category of representation-based models. It operates by calculating the embeddings of a query and a document, which involves averaging the word embeddings from their respective text fields. On the other hand, the interaction BERT-based model, referenced as \cite{b13}, employs a different approach. It concatenates the query and document text fields into a single sentence, which is then fed through multiple transformer layers. This approach has proven to yield state-of-the-art results, as cited in \cite{b14}. In recent few years, neural LTR models gradually start to be launched on various commercial search engines successfully, such as LinkedIn \citep{Guo:Liu:Detext:CIKM:2020}, Taobao \citep{Yao:Tan:Chen:Lear:2021}. 
\citet{DBLP:journals/corr/abs-2004-08476} introduced TFR-BERT, a generic document ranking framework that combines the power of LTR models and BERT.
\citet{DBLP:conf/www/WuMCPL022} proposed a multi-task learning approach for the ranking problem to optimize the different engagement signals simultaneously. 
Furthermore, some LTR models \cite{DBLP:conf/ictir/LuccheseNPT17, DBLP:conf/sigir/AslamKPSY09, DBLP:conf/sigir/KanoulasSMPA11, DBLP:conf/www/YangYCHLWXC20} used effective negative sampling technique to improve the efficiency of model training process and effectiveness of the resulting model by filtering out noise and reducing the redundancy of the query-document pairs. 
\citet{DBLP:conf/ictir/LuccheseNPT17} demonstrated the proportion of relevant documents to non-relevant documents highly affect the quality of the learning-to-rank collections. \citet{DBLP:conf/sigir/KanoulasSMPA11} illustrated the relevance grade distribution in the training set is an important factor for the effectiveness of learning to rank algorithms. 
While Neu-IR and traditional LTR models have shown promising results on large, well-labeled datasets, their performance decreases when faced with sparsely labeled data. Conversely, the global ranking function in LTR may not be optimal for document retrieval since it ignores differences between feature distributions for each query \cite{DBLP:conf/sigir/AiBGC18}. Although \citet{DBLP:conf/sigir/CanCM14, DBLP:conf/sigir/DehghaniZSKC17} constructed a ranking function specific to each query, this approach is limited by its high cost and low generalization ability due to the almost infinite number of queries and the unpredictable feature distribution of unseen queries. Our proposed MLTR model, however, can effectively improve the generalization of the LTR model on sparse datasets.

While traditional Learning to Rank (LTR) models rely on manually annotated labels for supervision, click-based LTR models harness user interaction logs as a guiding resource \cite{DBLP:conf/kdd/Joachims02}. Initially, these models were grounded in an online dueling-bandit framework \cite{ DBLP:conf/wsdm/SchuthOWR16, DBLP:conf/icml/YueJ09}, evolving later to an online pairwise method \cite{ DBLP:conf/cikm/OosterhuisR18} for unbiased pairwise optimization in LTR. However, that click data are not entirely reliable indicators of user preference due to various influencing factors beyond just preference \cite{DBLP:conf/sigir/JoachimsGPHG05, DBLP:conf/cikm/RadlinskiKJ08}. Addressing this, \citet{DBLP:journals/ijdsa/YuanGSLJ23} and later \citet{DBLP:conf/wsdm/JoachimsSS17} developed the first theoretically unbiased LTR method, based on the premise that a user’s likelihood of examining an item is tied to its rank position, with clicks primarily on items that are examined \cite{DBLP:conf/wsdm/CraswellZTR08, DBLP:conf/wsdm/WangGBMN18}. They employed counterfactual IPS estimation \cite{article_ROSENBAUM} to adjust for the biases inherent in these examination probabilities. IPS-based approaches, combating biases like position bias \cite{DBLP:conf/sigir/JoachimsGPHG05}, trust bias \cite{ DBLP:conf/www/AgarwalWLBN19}, and selection bias \cite{DBLP:conf/sigir/OosterhuisR20}, are prevalent in Information Retrieval (IR). \citet{DBLP:conf/ijcai/OosterhuisR21} enhanced these methods, accounting for potential changes in the logging policy during data collection. Despite their widespread use, IPS methods do grapple with high variance issues \cite{DBLP:conf/sigir/GuptaH0VO23}. This opens avenues for future research into alternative bias mitigation techniques, such as the doubly-robust method \cite{DBLP:journals/tois/Oosterhuis23}. Sample-based approximation \cite{DBLP:conf/sigir/Oosterhuis21, DBLP:conf/icml/UstimenkoP20} are typically employed in LTR scenarios where relevance is clearly established. Notably, \citet{DBLP:conf/sigir/Oosterhuis21} recently introduced the PL-Rank method, an efficient approach for unbiased gradient estimation based on sampled rankings. Click-based LTR models aim to derive unbiased ranking models from inherently biased user behavior data, demonstrating efficiency in unbiased learning to rank area. However, our research doesn't primarily focus on the bias in data collection or usage. Instead, we concentrate on the application of meta-learning techniques to train robust LTR-based models effectively, particularly in scenarios characterized by sparse data sets.

On the other hand, some works implement online LTR methods. For instance, \citet{DBLP:conf/icml/YueJ09} introduced the first online learning to rank method, which utilizes online evaluation by sampling model variants and comparing them with interleaving to identify better rankers, thereby improving the entire system. \citet{DBLP:conf/wsdm/HofmannSWR13} extended this by guiding exploration through reusing previous interactions. However, these approaches are not always efficient and sometimes their performance at convergence is much worse than offline approaches, especially in online settings where performance cannot be measured, making early-stopping unfeasible \cite{DBLP:conf/wsdm/SchuthOWR16}. Addressing these issues, our MLTR model can ensure more efficient optimization of traditional ranking model and keep stable model generalization with low computational cost in sparse data experimental settings.

\subsection{Sparsely Supervised Learning}

Supervised learning algorithms have faced a challenge in handling sparse and imbalanced labeled data. Previous techniques have typically addressed the negative impact of sparse data and imbalanced data distribution by optimizing the model or through data augmentation methods. From a model perspective, \citet{DBLP:conf/aaai/ZhouZY07} proposed that existing methods for dealing with the challenge of few labeled examples often rely on semi-supervised learning techniques that exploit both labeled and unlabeled data. Moreover, \citet{DBLP:journals/ijon/SunH10} introduced an active learning strategy for identifying informative examples that require manual labeling, which is particularly beneficial when manual labeling resources are limited. Nonetheless, it is crucial to note that the inductive bias of a model can be significantly impacted by having a limited number of examples, commonly referred to as sparse data, as noted in \cite{DBLP:journals/jair/Baxter00}. From data augmentation perspective, resampling is an typical technique for handling data imbalance in machine learning \cite{10.5555/1211118}. Data oversampling was introduced by \citet{chawla2003c4}, who sampled the minor classes from the available data and included them in the training process to mitigate the imbalance between major and minor classes. One of the popular oversampling techniques is SMOTE \cite{DBLP:journals/jair/ChawlaBHK02}, which has various adaptations such as those proposed by \cite{DBLP:conf/icic/HanWM05, DBLP:conf/ijcnn/HeBGL08}, and others. However, the learned supervised model has limit improvement with duplicated data without new information and high risk of overfitting. In the same way, \citet{DBLP:journals/tsmc/LiuWZ09} employed data undersampling as a technique to achieve a comparable amount of training data in various classes by reducing the number of data in the major classes. The article referenced as \cite{DBLP:conf/www/YangYCHLWXC20} reports on the use of a two-tower neural model that was trained utilizing a mixed negative sampling technique alongside batch random negatives. However, this method may lead to a loss of information during the reduction of training data through sampling. Data generation models for informative data augmentation in LTR are proposed by \citet{DBLP:conf/aaai/YuL19} and \citet{DBLP:conf/cikm/QiuJCZ21}, as they believe that generating informative data is more beneficial than using resampling techniques. Those models generated informative synthetic data based on Adversarial Autoencoder (AAE) \cite{DBLP:journals/corr/MakhzaniSJG15} and Gaussian Mixture Variational Autoencoder (GMVAE) \cite{DBLP:journals/corr/DilokthanakulMG16}, respectively. Given the strong text generation capabilities of large language models (LLMs), many researchers \cite{DBLP:journals/corr/abs-2202-05144, DBLP:conf/iclr/DaiZMLNLBGHC23, DBLP:journals/corr/abs-2307-08303} propose using LLM-driven methods to generate pseudo queries or relevance labels from existing collections. Both of them could generate new data given different query types and different relevance levels. Resampling methods and data augmentation techniques have the potential to mitigate the effects of imbalanced data in the training set, however, they have limited improvement on overall model generalization.

\subsection{Meta-Learning for Information Retrieval}

Meta-learning is also known as learning to learn, which aims to learn better algorithms, including better parameter initialization, optimization strategy \cite{DBLP:conf/nips/AndrychowiczDCH16}, network architecture \cite{DBLP:conf/iclr/ZophL17} and distance metrics \cite{DBLP:conf/emnlp/GaoHZLLSZ19}. \citet{DBLP:conf/icml/FinnAL17} proposed a Model-Agnostic Meta-Learning (MAML) algorithm, which trains a model on a variety of tasks, such that the model can be easily generalized to a new task with a small number of gradient steps from a small number of data from that task. Also, a lot of existing works have implemented the meta-learning approach in other research areas. \citet{DBLP:conf/kdd/LeeIJCC19} proposed Meta-Learned User Preference Estimator (MeLU), which utilizes meta-learning approach to deal with the cold start problem in the recommendation system. \citet{DBLP:journals/tois/CuiS0Y022} proposed a novel approach to address the challenge of data sparsity in next POI recommendation, called Meta-SKR, which leverages a meta-learning approach to generate user-conditioned parameters for a sequential-knowledge-aware embedding module. \citet{DBLP:conf/coling/BansalJM20} proposed a MAML-based meta-learning method LEOPARD
for domain adaptation tasks in NLP. 
In addition, there are some works on information retrieval. 
\citet{carvalho2008meta} proposed a meta-learning algorithm to suppress the undesirable outlier effects of the pairs of documents using the pairwise ranking function. \citet{DBLP:conf/pkdd/ZabashtaSF15} presented a meta-learning model for selecting rank aggregation algorithms based on a specific optimality criterion. \citet{DBLP:conf/www/WuMZL22} introduced a novel Bayesian Online Meta-Learning Model (BOML) tailored for personalized product search. BOML harnesses meta-knowledge acquired from inferences made about other users' preferences, enabling accurate predictions even in situations where historical data is limited. By addressing the challenge of data sparsity, BOML can significantly enhance the accuracy of recommendations in personalized product search.
\citet{DBLP:conf/sigir/WangT022} proposed Meta-learning based Fair Ranking (MFR), which alleviates the data bias and achieves better fairness metrics in the ranking model through an automatically weighted loss. 
\citet{DBLP:conf/acl/SunQLXZBLB20} proposed the MetaAdaptRank, which is a domain adaptive learning method for few-shot Neu-IR based on meta-reweighted weak supervision data selection during the different periods of the training process. However, to the best of our knowledge, there have been no work using meta-learning on ranking models with the sparsely labeled queries.

\section{The Framework}

\begin{table}
  % \begin{center}
  \caption{Summary of notation.}
    \label{tab:notation}
    \begin{tabular}{p{0.12\textwidth}|p{0.74\textwidth}}
      \toprule
      \bf Notation & \bf Definition \\
      \hline
      $q$, $d$ & query, document\\
      $\mathcal{S}$ & set of datasets with queries for meta training\\
      $\mathcal{S}_{train}$ & set of sampled training data for meta training $\mathcal{S}$ for query-specific ranker\\
      $\mathcal{S}_{test}$ & set of sampled test data from meta training $\mathcal{S}$ for meta ranker \\
      $\mathcal{S}_{train: p\cdot n\cdot}$ & set of fixed positively and negatively labeled items assigned to each query in the training dataset of $\mathcal{S}_{train}$\\
      $\mathcal{S}_{test: p\cdot n\cdot}$ &  set of fixed positively and negatively labeled items assigned to each query in the test dataset of $\mathcal{S}_{test}$\\
      $\mathcal{S}_{train, i}$& training data for query $i$ in $\mathcal{S}_{train}$\\
      $\mathcal{S}_{test, i}$& test data for query $i$ in $\mathcal{S}_{test}$\\
      $\mathcal{T}$& set of datasets with unseen queries for fine-tuning and evaluation\\
      $\mathcal{T}_{tuning}$ & set of sampled data from $\mathcal{T}$ for fine-tuning \\
      $\mathcal{T}_{eval}$& set of sampled data from $\mathcal{T}$ for evaluation\\
      $\mathcal{T}_{tuning: p\cdot n\cdot}$ & set of fixed positively and negatively labeled items assigned to each query in the test data $\mathcal{T}$ for fine-tuning\\
      $\mathcal{T}_{eval: rest}$ & remaining test data set $\mathcal{T}$ for evaluation \\
      $\mathcal{T}_{tuning, i}$& fine-tuning data for query $i$ in $\mathcal{T}_{tuning}$\\
      $\mathcal{T}_{eval, i}$& evaluation data for query $i$ in $\mathcal{T}_{eval}$\\
      $g(\cdot, \theta)$ & meta ranker with global parameter $\theta$\\
      $f(\cdot, \theta_i)$ & query-specific ranker with parameter $\theta_i$ for query $i$\\
      $\mathcal{L}_{query}(\theta_i)$ & loss function of query-specific ranker \\
      $\mathcal{L}_{meta}(\theta)$ & loss function of meta ranker \\
      \bottomrule
    \end{tabular}
    
  % \end{center}
\end{table}

Our proposed Meta Learning to Rank (MLTR) framework is presented in this section. First, we will explain the traditional LTR model and the meta-based LTR model, which sets the problem context. Then, we will detail the MLTR's training and testing processes, which enables fast adaptation and improve model generalization.

\subsection{Learning to Rank}

Let $Q=\{q_1, q_2,..., q_N\}$ denote the collection of $N$ queries, $D=\{d_1, d_2,..., d_M\}$ denote the collection of $M$ documents, and $L=\{1, 2,...,l\}$ denote the collection of $l$ labels. There has an order of labels $l > l-1 > ... > 1$, where $>$ denote the sequence of the label order.

For every query $q_i$, there is a corresponding related document collection $D_i = \{d_{i,1}, d_{i,2}, ..., d_{i,J}\}$ and the corresponding label collection $y_i = \{y_{i,1}, y_{i,2}, ..., y_{i,J}\}$. Above all, the original data $\mathcal{S}$ can be denoted as $\mathcal{S} = \{(q_i, D_i), y_i\}_{i=1}^N$. The object is to train the ranking model of a given query $q_i$ and corresponding related document collection $D_i$ with the ranking label $y_i$, mathematically as $\hat{y}_{i}=f(\mathbf{x}_{i}; \theta)$ where $f(\cdot)$ is a ranking function, $\theta$ represent all the learnable parameters in $f(\cdot)$ and $\mathbf{x}_i$ denote the concatenated feature vector generated from the query and documents $(q_i, D_i)$. $\mathbf{x}_i = concat(\phi(q_i), \psi(D_i), \mathbf{r}_i)$, where $\phi(\cdot)$ and $\psi(\cdot)$ denote the query encoder and the document encoder respectively; $\mathbf{r}_i$ denotes the numeric ranking features for each query and corresponding related document collection $(q_i, D_i)$. Generally, we learn the optimized $\theta^\ast$ by $min_\theta \frac{1}{N} \sum_{i=1}^N\mathcal{L}(\hat{y}_i, y_i)$ and $\mathcal{L}$ could be used as any ranking loss functions.

\subsection{Problem Formulation}

Our work is inspired by optimization-based meta-learning, specifically Model-Agnostic Meta-Learning (MAML) \cite{DBLP:conf/icml/FinnAL17}, which optimizes globally shared parameters over several tasks, so as to rapidly adapt to a new task with just one or a few gradient steps based on a small number of examples.

In the search and ranking setting, we define each task as ranking items for a given query. Our MLTR framework trains a model with a good generalization which can quickly adapt to a new query based on the query's sparse engagement information. We divide the raw data into $\mathcal{S}$ and $\mathcal{T}$. We limit each query task (including the query and all its corresponding items) within only one set, such that there is no query overlap between $\mathcal{S}$ and $\mathcal{T}$. 

For each task query $q_i$ ($\in Q$) in $\mathcal{S}$, its corresponding items are randomly divided into a training set $\mathcal{S}_{train, i}$ and a test set $\mathcal{S}_{test, i}$ to optimize the model during various stages.
For each task query $q_i$ ($\in Q$) in $\mathcal{T}$, its corresponding items are randomly split into a fine-tuning set $\mathcal{S}_{tuning, i}$ and an evaluation set $\mathcal{S}_{eval, i}$ to fine-tune the model and assess its performance, respectively. For further information regarding the notation employed in this paper, please refer to Table \ref{tab:notation}.

\subsection{Meta Learning to Rank}

\begin{figure}[t]
\centering
\includegraphics[width=0.95\linewidth]{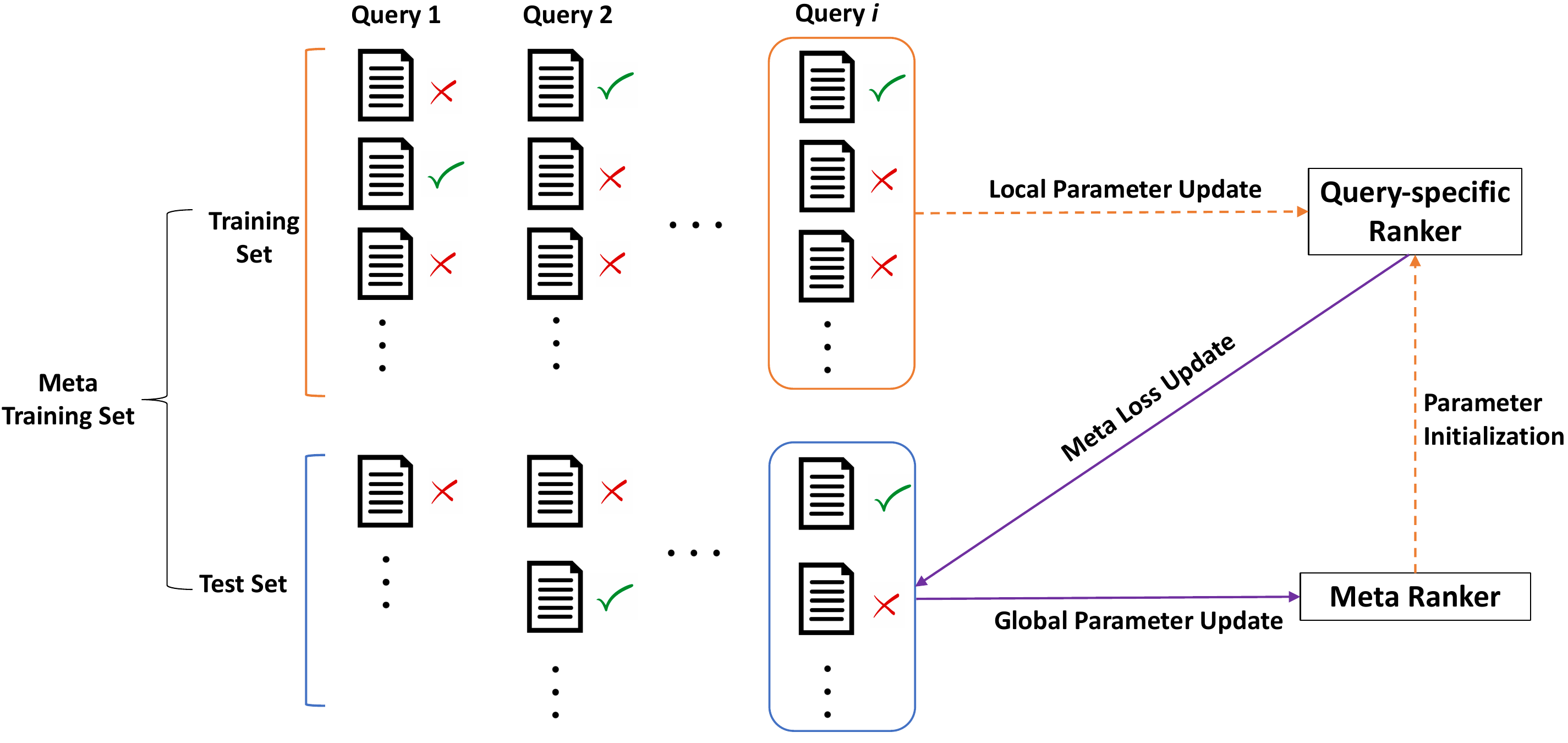}
\caption{The architecture of the proposed meta learning to rank framework (MLTR) in the meta-training process}
\label{Fig.framework} 
\end{figure}

The MLTR framework's key concept is to create robust model parameters through many query-based ranking tasks in meta-training, then quickly adapt these parameters for new tasks in meta-testing with a few gradient steps. In meta-training, it performs local and global updates. The local update adjusts the parameter of the query-specific ranker on each training set (learning process). The global update trains the parameter of the global ranker to minimize the meta losses with the adapted parameters on the test sets (learning-to-learn process). The proposed meta-learning approach to ranking considers that individual queries may have distinct optimal parameters for their rankers, which is unlike traditional Learning to Rank (LTR) models that learn a global ranking model applicable to all queries. Fig. \ref{Fig.framework} illustrates the architecture of the proposed meta-training process. To estimate document relevance for a new query in meta-testing, the ranker can then be fine-tuned based on the limited amount of labeled documents. The following subsections provide the details of the proposed approach.

\subsubsection{Meta Training} 

We define the meta ranker and query-specific ranker in our model as well, similar as the MAML \cite{DBLP:conf/icml/FinnAL17} setting. The ranker model can be defined with any model structure based on your tasks, such as the basic multi-layer perceptron (MLP). Query-specific ranker $f(\cdot; \theta_i)$ is initialized by the meta ranker and learns the task-specific parameters $\theta_i$ to optimize a specific task at a time. Meta ranker $g(\cdot; \theta)$ learns across multiple tasks based on the query-specific ranker, and can improve the model generalization performance. Although these two rankers share the same network structure and parameters $\theta_i$, $\theta$, respectively, their loss function objectives are different. Thus, the meta-learning for sparsely supervised search could be defined as, for the meta training dataset $\mathcal{S} = \mathcal{S}_{train} \cup \mathcal{S}_{test}$, the meta-train process aims to train the query-specific ranker $f(\cdot; \theta_i)$ to learn task-specific parameters $\theta_i$ on $\mathcal{S}_{train}$, and to train the meta ranker $g(\cdot; \theta)$ cross multiple tasks on $\mathcal{S}_{test}$ to extend the model generalization. Note that training and test sets are split at the query level within meta training process. As introduced earlier in this section, each query $q_i$ in the meta-train data has a corresponding training set $\mathcal{S}_{train, i}$ ($\subset \mathcal{S}_{train}$) and $\mathcal{S}_{test, i}$ ($\subset \mathcal{S}_{test}$).

The model training consists of a basic specific-task learning process and cross-task meta adaptation process,
trained on training set $\mathcal{S}_{train}$ and test set $\mathcal{S}_{test}$, respectively.
Note that there is no intersection between $\mathcal{S}_{train}$ and $\mathcal{S}_{test}$.
For the basic specific-task learning process (local parameter updates with training set), 
the query-specific ranker focuses on the quick acquisition of knowledge to learn task-specific parameters through the LTR loss. LTR loss indicates how well the model is performing on the specific task (query). 
For the meta cross-task adaptation process (global updates with test set), the model further learns generalized parameters cross-tasks and updates the meta ranker through the meta loss. Meta loss indicates how well the model is performing across multiple tasks. In attempting to learn a meta ranker this way, it could solve the generalization issue, especially in the sparsely labeled dataset.

\begin{figure}[htb]
  \centering
  \begin{minipage}{.95\linewidth}
    \begin{algorithm}[H]
    % \begin{algorithm}[tb]
    \caption{Meta Learning to Rank (MLTR)} 
	\label{alg1} 
	\begin{algorithmic}[1]
		\REQUIRE $p(\mathcal{S})$: distribution over query-level tasks
            \REQUIRE $\alpha, \beta$: step size hyperparameters, $K$: sampled items number, $T$: inner loop number
		\STATE Randomly initialize $\theta$ for meta ranker $g(\cdot)$
            \WHILE{not done}
                \STATE Sample a batch of queries $\mathcal{S}_B$ from $p(\mathcal{S})$
                \FOR{each query $q_i \in \mathcal{S}_B$}
                    \STATE Initialize query-specific ranker parameters $\theta_i = \theta$
                    \FOR{inner loop $t=1, \ldots, T$}
                        \STATE Sample $K$ items $\mathcal{S}_{train: K, i}$ from $Q_i$ based on a sample strategy
                        \STATE Evaluate $\nabla \mathcal{L}_{query}(\theta_i)$ using $\mathcal{S}_{train: K, i}$ and $\mathcal{L}_{query}(\theta_i)$ in Equation~(\ref{eq:baseloss})
                        \STATE Compute query-specific ranker parameters $\theta_i$ with gradient descent in Equation~(\ref{eq:baseupdate})
                    \ENDFOR
                    \STATE Sample $K$ items $\mathcal{S}_{test: K, i}$ from $Q_i$ based on a sample strategy
                    \STATE Add $\mathcal{S}_{test: K, i}$ to $\mathcal{S}_{B: test}$
                \ENDFOR
                \STATE Evaluate $\nabla \mathcal{L}_{meta}(\theta)$ using $\mathcal{S}_{B: test}$ and $\mathcal{L}_{meta}(\theta)$ in Equation~(\ref{eq:metaloss})
                \STATE Update meta ranker $\theta$ in Equation~(\ref{eq:metaupdate})
            \ENDWHILE
    \end{algorithmic} 
\end{algorithm}
\end{minipage}
\end{figure}

Algorithm \ref{alg1} shows the detailed steps of the meta training process.
First, we define two different learning rates $\alpha$ and $\beta$ for query-specific ranker parameter updates and meta ranker parameter updates, respectively. The model starts with initializing the meta ranker parameters. Then it updates the parameters based on each batch, until convergence. For each batch, meta training process could be summarized as following steps: first, initialize the query-specific ranker $f(\cdot; \theta_i)$ with the meta ranker $g(\cdot; \theta)$ parameters $\theta_i = \theta$. Secondly, sample a batch of queries $\mathcal{S}_B$ from the $\mathcal{S}$, $B$ denotes the batch size. Then, we can rewrite the loss function of query-specific ranker $\mathcal{L}_{query}$ for each query as the following: 
\begin{equation}
\label{eq:baseloss}
    \begin{aligned}
    \mathcal{L}_{query}(\theta_i) = \mathcal{L}_{\mathcal{S}_{train, i}}(\hat{y}_i, y_i) = \mathcal{L}_{\mathcal{S}_{train, i}}(f(\mathbf{x}_i; \theta_i), y_i)
    \end{aligned}
\end{equation}
where $\mathcal{S}_{train, i}$ represents the training set of query $q_i \in \mathcal{S}_B$, and $\hat{y}_i = f(\mathbf{x}_i; \theta_i)$ represents the model output of query $q_i$. $\mathcal{L}$ denotes the different ranking loss. This query-specific ranker aims to find optimal parameters $\theta_i$ for query $q_i$. It will be updated sequentially multiple times (denoted by $T$) through an inner loop. For each step in the inner training, $K$ items are sampled from $q_i$'s document collection $D_i$ as the training set for this step.

Next, the query-specific ranker parameters $\theta_i$ are updated by gradient descent of the query-specific loss $\mathcal{L}_{query}$ as the following:
\begin{equation}
\label{eq:baseupdate}
    \begin{aligned}
        \theta_i = \theta_i - \alpha\nabla \mathcal{L}_{query}(\theta_i)
    \end{aligned}
\end{equation}
where $\alpha$ denotes the learning rate of query-specific ranker $f(\cdot; \theta_i)$.

After updating the query-specific ranker $f(\cdot; \theta_i)$ for the task associated with query $q_i$, we sample $K$ items from $D_i$ to form the test set. This sampling excludes items from the training set used in the last inner loop $T$. It is important to note that an item may be present in the test set of the last inner loop $T$ and also in the training sets of earlier inner loops $1, \ldots, T-1$. However, this overlap does not lead to data leakage issues, as both rankers operate within the scope of the meta-training process. Next step, we need to calculate the meta loss and update the meta ranker's parameters for optimizing all query-based tasks within batch $\mathcal{S}_B$. The meta loss $\mathcal{L}_{meta}$ will be defined as
\begin{equation}
\label{eq:metaloss}
    \begin{aligned}
        \mathcal{L}_{meta}(\theta) = \mathcal{L}_{\mathcal{S}_{B: test}}(\hat{y}_i, y_i) = \frac{1}{B} \sum_{i=1}^B\mathcal{L}_{\mathcal{S}_{test, i}}(g(\mathbf{x}_i; \theta_i), y_i)
    \end{aligned}
\end{equation}
where $\mathcal{S}_{B: test}$ represents the test set of query batch $S_B$, and $\hat{y}_i = g(\mathbf{x}_i; \theta_i)$ represents the model output of each query $q_i$ respect to the meta ranker $g(\cdot, \theta)$ and the updated parameters $\theta_i$ from the query-specific ranker, based on the training set. We let $\mathcal{L}$ denotes the different ranking loss. We sum up the loss from each query of batch $\mathcal{S}_B$ and compute the average loss as the meta loss from this batch. This meta ranker aims to optimize parameters $\theta$ through the batch query $\mathcal{S}_B$, learns across multiple query-level tasks based on the query-specific ranker, and can improve the model generalization performance. $K$ items are sampled from $q_i$'s document collection $D_i$ as the test set for this step.

Meta ranker updates $\theta$ by gradient descent of the meta loss $\mathcal{L}_{meta}$ as the following:
\begin{equation}
\label{eq:metaupdate}
    \begin{aligned}
        \theta = \theta - \beta \nabla \mathcal{L}_{meta}(\theta)
    \end{aligned}
\end{equation}
where $\beta$ denotes the learning rate of meta ranker $g(\cdot; \theta)$.

\begin{figure}[t]
\centering
\includegraphics[width=0.8\linewidth]{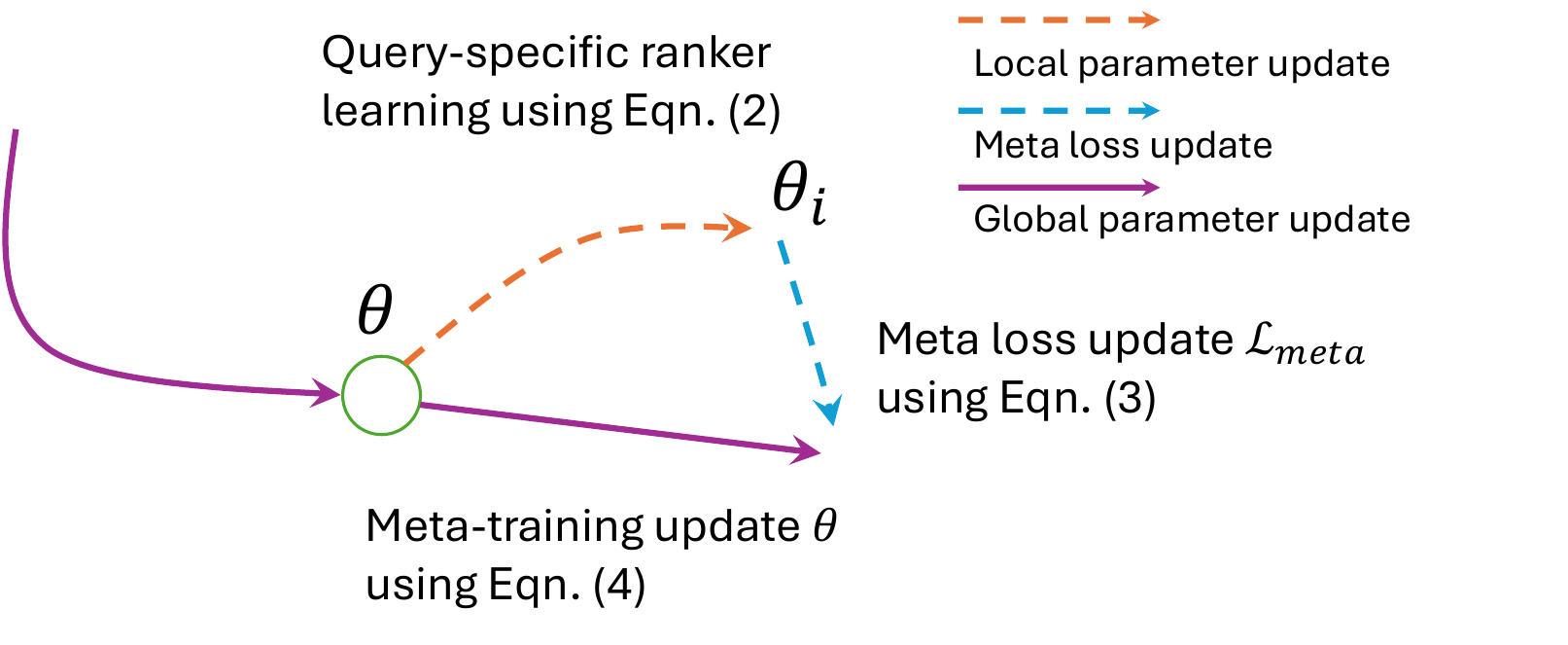}
\caption{Illustration of MLTR in meta training which optimizes for a representation $\theta$ that can quickly adapt to new queries. The orange dashed line represents the query-specific ranker initialized from the meta ranker and locally updated based on the training set data in meta training. The blue dashed line represents the direction of meta loss updates based on the updated query-specific ranker on test data in meta training. The purple solid line represents the global updates of the meta ranker based on the meta loss.}
\label{Fig.update} 
\end{figure}

Repeating the above batch level meta training process, the query-specific ranker will continuously train on the training set $\mathcal{S}_{train}$, the meta ranker will adapt and update meta-parameters $\theta$ on the test set $\mathcal{S}_{test}$ until the model parameter converges. Fig. \ref{Fig.update} shows an illustration of MLTR in meta training which optimizes for a representation $\theta$ that can quickly adapt to new queries.

\subsubsection{Meta Testing} 
During the meta testing phase,
the meta-trained model (meta ranker $g(\cdot)$) is used to make predictions on the meta-test queries/tasks.
Different from the usual supervised learning model, the meta-trained model has a further fine-tuning process for additional gradient steps with few epochs on the fine-tuning set $\mathcal{T}_{tuning}$ before running the inference. 
This additional fine-tuning step enables the model parameters to fast adapt to new queries based on the Equation~(\ref{eq:baseupdate}), due to the learning-to-learn process. This meta testing process accounts the fact that different queries have different optimal parameters for their rankers and thus reduce the impact of domain shift on the model. In consequence, the MLTR model has a significant advantage, particularly when facing new queries with distinct characteristics compared to the queries used in training. Additionally, we can also disable the fine-tuning mechanism and evaluate the model's performance using the evaluation dataset. The experiments in Section 5 demonstrate that MLTR with or without fine-tuning both improves model generalization for sparsely supervised queries.

\section{Experimental Setup}

\subsection{Datasets}

\label{sec:dataset}
We evaluate the performance of our MLTR framework in the setting of sparsely supervised queries using four different datasets. The datasets include MQ2007, MQ2008, MSLR-10K \footnote{\url{https://www.microsoft.com/en-us/research/project/letor-learning-rank-information-retrieval}} and Istella-S LETOR \footnote{\url{http://quickrank.isti.cnr.it/istella-dataset}}, which are public datasets widely used as benchmarks for LTR models \cite{Qin2013IntroducingL4}. These datasets consist of queries, retrieved documents, and labels provided by human experts. Furthermore, we used a real-world e-commerce dataset collected from a one-month user log on Walmart.com. The focus of the dataset is on non-frequent tail queries, meaning the label distribution is extremely sparse. This dataset includes user search queries and the corresponding products in the search results, with labels (rating scores) ranging from 1 to 15 based on the level of user engagement.
Query-product pairs that have been purchased receive the highest scores, whereas products that have received only clicks are assigned scores ranging in the middle. Products that have only received impressions are assigned scores lower than that of the click-only products. Negative items are assigned a score of 1. Scores for ordered products are calculated using a smoothed estimation of their order rate ($rate = \frac{order + \alpha }{impressions + \alpha}$), where $\alpha$ is the smoothing factor. 

For all the above datasets, we first divide the raw data into meta-train, meta-validation, and meta-test sets, with a ratio of 80\%, 10\%, and 10\%, respectively. Each query-document (item) pair is associated with a relevant rating label, which has different ranges for each dataset. Table \ref{tb:staticstic} provides more details about the data statistics.

\begin{table}[t]
\centering
\caption{Basic statistics of the datasets}
\resizebox{\linewidth}{!}{
\begin{tabular}{ccccccc}
\toprule
 & \bf Queries & \bf Items & \bf Query-Item pairs & \bf Positives & \bf Features & \bf Range of ratings \\
\hline
MQ2007 & 1,692 & 65,323 & 69,623 & 25.84\% & Sparse features (46) & 0$\sim$2\\
MQ2008 & 784 & 14,384 & 15,211 & 19.28\% & Sparse features (46) & 0$\sim$2\\
MSLR-10k & 10,000 & 1,200,192 & 1,200,192 & 47.99\% & Sparse features (136) & 0$\sim$4\\
Istella-S & 33,018 & 3,408,630 & 3,408,630 & 11.39\% & Sparse features (220) & 0$\sim$4\\
Walmart Dataset & 151,770 & 12,372,081 & 38,837,815 & 2.19\% & Re-ranking feature (63), text feature & 1$\sim$15\\
\bottomrule
\end{tabular}
}
\label{tb:staticstic}
\end{table}

\subsection{Sparsely Labeled Data}

\label{sample_data}
To simulate the sparse labeled queries, we further process the train, validation, and test datasets. In our experiments, we primarily control the number of positively labeled documents since it is usually limited in the real world and the negative documents can often be sampled from the dataset in a relatively large quantity. We perform a quantitative comparison on the simulated imbalanced datasets during training and testing. 

We use $\mathcal{S}$ with superscript $p \cdot n \cdot$ to denote the number of sampled positive/negative labeled items per query in the training data (e.g., $\mathcal{S}_{train:p1n9}$ means the sampled training data with 1 positive-labeled items and 9 negative-labeled items per query). Thus, $\mathcal{S}_{train}$ is chosen from 
$$\{\mathcal{S}_{train: p1n4}, \mathcal{S}_{train: p1n9}, \mathcal{S}_{train: p1n19}, \mathcal{S}_{train: p1n29}, \mathcal{S}_{train: p1n39} \}$$
and $\mathcal{S}_{test}$ is chosen from 
$$\{\mathcal{S}_{test: p1n4}, \mathcal{S}_{test: p1n9}, \mathcal{S}_{test: p1n19}, \mathcal{S}_{test: p1n29}, \mathcal{S}_{test: p1n39} \}$$

Similarly, we use $\mathcal{T}$ with superscript $p \cdot n \cdot$ to denote the number of sampled positive/negative labeled items per query for model fine-tuning, $\mathcal{T}_{tuning}$ is chosen from
$$\{\mathcal{T}_{tuning:p1n4}, \mathcal{T}_{tuning:p1n9}, \mathcal{T}_{tuning:p1n19}, \mathcal{T}_{tuning:p1n29}, \mathcal{T}_{tuning:p1n39} \}$$

and the rest of the items of each query to evaluation our model with $\mathcal{T}_{eval:rest}$. 

The validation set will be used to find the best model and hyper-parameters during the meta-training process, and will be split in the same manner as the meta-test dataset.

\subsection{Evaluation Metrics}

% We aim to evaluate the ranking results of the proposed work. 
For the evaluation of the ranking results in MLTR, we apply Normalized Discounted Cumulative Gain (NDCG) which is suitable for ranking where users are usually sensitive to the ranked position of the relevant items \cite{DBLP:journals/tois/JarvelinK02}.

\subsection{Baseline Methods}

To verify the efficiency and compatibility of our proposed model, we refrained from directly using overly complex baseline models. Instead, we conducted experiments on simple models and observed the resulting improvements to verify the effectiveness of our meta-learning based method for overall performance improvement in LTR models. On the other hand, due to the lack of semantic features for queries and corresponding documents in the public datasets MQ2007, MQ2008, MLSR-10K, and Istella-S, we did not utilize the corresponding text embedding representation features in our tests. Nevertheless, we supplemented the corresponding text embedding representation features using the BERT model with pre-trained weights from distillbert-base-uncased\footnote{\url{https://huggingface.co/distilbert-base-uncased}} based on the text information of the query and document in the subsequent Walmart.com dataset. We then conducted experiments to verify the effectiveness of these features. We compare MLTR with the following competitive baselines:

\begin{itemize} 
\item \textbf{LTR}: The LTR baseline is a 3-layer Multi-Layer Perceptron (MLP) with a ReLu activation function. The ranking loss functions are introduced later in this section. To ensure fair comparisons, we perform fine-tuning on the test stage.
\item \textbf{LTR+SMOTE} \cite{DBLP:journals/jair/ChawlaBHK02}: This method is resampling-based and generates a resampled list using SMOTE, a popular oversampling strategy. We added the resampled data to the original training data and followed the same training and testing protocol as the LTR baseline model.
\item \textbf{LTR+GMVAE} \cite{DBLP:conf/cikm/QiuJCZ21}: This method is based on data augmentation and utilizes GMVAE to generate additional synthetic items. The GMVAE model is pre-trained with the entire training dataset, and then the augmented lists are produced. We added the synthetic data to the original training data and followed the same training and testing approach used with the LTR baseline model.
\item \textbf{LTR+Policy-Gradient} \cite{DBLP:conf/sigir/Oosterhuis22}: Plackett-Luce (PL) ranking models, a decision theory-based approach to ranking. This model employs Gumbel sampling techniques for efficient sampling of multiple rankings from a PL model. Following this, algorithms PL-Rank-1, PL-Rank-2, and PL-Rank-3 are applied to these samples. This process enables an unbiased approximation of the gradient of a ranking metric in relation to the model. For our experiments, we have adhered to the model parameters and implementation as detailed in the official repository \footnote{\url{https://github.com/HarrieO/2022-SIGIR-plackett-luce}} and adapted the data-loader to fit our experimental setup.
\item \textbf{LTR+Unbiased} \cite{DBLP:journals/tois/Oosterhuis23}: Unbiased click-based LTR models, tailored specifically to adjust for position bias in click feedback. In our experiments, we deploy three distinct estimators. First, the inverse-propensity-scoring (IPS) approach employs counterfactual IPS estimation to mitigate the selection bias linked to examination probabilities. Next, we utilize DM and DR approaches that account for position bias, trust bias, and item-selection bias, offering a more flexible criterion for unbiasedness compared to the widely used IPS method. Our implementation follows the model parameters outlined in the official repository \footnote{\url{https://github.com/HarrieO/2022-doubly-robust-LTR}} with $N=10^{-4}$ for comparison, and we have adapted the data-loader to suit our experimental framework.

\end{itemize}
To ensure a fair comparison, most components in our proposed MLTR model employ the same model structure as the LTR baseline model. Regarding MLTR, we made corresponding adjustments to the model training process and data usage.

In traditional LTR models, three different types of loss functions, namely Pointwise, Pairwise, and Listwise \cite{liu2009learning}, are usually used depending on the task and data.
We use the following representative losses for the LTR baseline and MLTR: RankMSE, RankNet, LambdaRank, and ListNet losses. 

\textbf{Pointwise loss}: It only takes into account a single document $d_{i,j}$ at a time for a query $q_i$. RankMSE algorithm \cite{DBLP:journals/tit/CossockZ08} is as follows,
\begin{equation}
\label{eq:pointwise}
    \mathcal{L}(f;\mathbf{x}_{i,j}, y_{i,j}) = \sum_{j=1}^{M}(f(\mathbf{x}_{i,j}) - y_{i,j})^2
\end{equation}

\textbf{Pairwise loss}: It considers a pair of documents $<d_{i,j}, d_{i,s}>$ at a time for a query $q_i$ if $y_{i,j} > y_{i, s}$ ($d_{i,j}$ should be ranked before $d_{i,s}$) \citep{burges2010from}. 
RankNet algorithm \citep{DBLP:conf/icml/BurgesSRLDHH05}
and LambdaRank algorithm \citep{Wang:Li:Lambdaloss:CIKM:2018}, with their loss functions shown 
in Equation~(\ref{eq:pairwise}) as follows:

\begin{equation}
    \label{eq:pairwise}
    \mathcal{L}(f;\mathbf{x}_{i,j}, \mathbf{x}_{i,s}) = 
    \sum_{j=1}^{M-1}\sum_{s=1, y_{i,j}>y_{i,s}}^M\varphi(f(\mathbf{x}_{i,j}) - f(\mathbf{x}_{i,s}))
\end{equation}
where $\varphi$ denotes the Sigmoid function for RankNet loss, $\varphi(u)= \Delta NDCG(j,s) \log_2(1+ e^{-\sigma u})$ for LambdaRank loss, where $\sigma$ is a hyper-parameter and $\Delta NDCG(j,s)$ is the absolute difference between the NDCG values when two documents $d_{i,j}$ and $d_{i,s}$ are swapped.

\textbf{Listwise loss}:  It directly looks at the entire list of documents $D_i$ and tries to come up with the optimal ordering for each query $q_i$ \citep{DBLP:conf/icml/CaoQLTL07}. For example, the loss function for the ListNet algorithm is as follows,
\begin{equation}
    \label{eq:listwise}
    \mathcal{L}(f;\mathbf{x}_i,y_i) = \sum_{i=1}^N{L}(f(\mathbf{x}_i), y_i)
\end{equation}
where $L(\cdot)$ denote the cross-entropy loss. $f(\mathbf{x}_i)$ is the predict label for query $q_i$. $y_i$ denote the true label of each document in query $q_i$.

\subsection{Research Questions}
An extensive set of experiments were designed to address
the following questions of the proposed research:

\textbf{RQ1}: Can the proposed MLTR framework achieve improved performance on sparsely labeled queries over the baseline methods? (Section~\ref{sec:experi_baseline})

\textbf{RQ2}: How does the training and test mechanism designed for MLTR effectively improve model performance compared to traditional model training processes? (Section~\ref{sec:meta_train_test})

\textbf{RQ3}: Without fine-tuning towards a specific query in test data, can MLTR still improve model generalization? (Section~\ref{sec:without_fine_tuning} and \ref{sec:MLTR_qd_pair})

\textbf{RQ4}: Can MLTR alleviate the data sparsity issue and domain shift problem? How much NDCG lift can the MLTR models gain over the traditional LTR models? Is the amount of NDCG relative gain correlated with training/test data's sparseness? (Section~\ref{sec:robustness})

\textbf{RQ5}: Can MLTR be effective in real-world applications with limited labeled data and result in improved performance? (Section~\ref{sec:case_study})

\section{Experimental Results}
\label{sec:experi}

In this section, we conduct experiments on the datasets introduced in Section~\ref{sec:dataset}. 
We compare the proposed MLTR model and the baseline LTR models under different scenarios, taking into account
multiple ranking loss functions and multiple simulated data sparsity cases. 

\subsection{Baseline Comparison (RQ1)}

\begin{table}[t]
\centering
\caption{Comparative Performance of Baseline Models and the MLTR Framework in Terms of NDCG@1, NDCG@5, and NDCG@10 Metrics on the Evaluation Set $\mathcal{T}_{eval:rest}$. This evaluation encompasses four publicly available datasets: MQ2007, MQ2008, MSLR-10K, and Istella-S. The highest-scoring results for each task and metric are emphasized. The symbol $\ddagger$ indicates a statistically significant improvement of MLTR (with and without fine-tuning) over the corresponding LTR models. This is evidenced by a p-value $< 0.01$ in a two-tailed t-test.}
\resizebox{\textwidth}{!}{%
\begin{tabular}{ccccc}
\toprule
\multicolumn{1}{c}{} & \bf MQ2007 & \bf MQ2008 & \bf MSLR-10K & \bf Istella-S\\
\bf Method &\bf NDCG@1 / 5 / 10 & \bf NDCG@1/ 5 / 10 &\bf NDCG@1 / 5 / 10 &\bf NDCG@1 / 5 / 10\\
\hline
\multicolumn{5}{c}{\multirow{2}{*}{\textbf{LTR}}}  \\
\multicolumn{5}{c}{} \\
RankNet & 0.4090 / 0.4672 / 0.5166 & 0.5000 / 0.5931 / 0.6461 & 0.3392 / 0.3594 / 0.3886 & 0.6146 / 0.6335 / 0.702  \\ 
RankMSE & 0.4286 / 0.4501 / 0.4946 & 0.4583 / 0.5407 / 0.6030 & 0.3624 / 0.3723 / 0.3974 & 0.6255 / 0.6412 / 0.706 \\
ListNet & 0.4454 / 0.4857 / 0.5354 & 0.4722 / 0.5796 / 0.6309 & 0.3984 / 0.3994 / 0.4224 & 0.6279 / 0.6433 / 0.7088 \\
LambdaRank & 0.4314 / 0.5025 / 0.5441 & 0.5972 / 0.6264 / 0.6907 & 0.3731 / 0.3808 / 0.4097 & 0.6146 / 0.6354 / 0.7054 \\
\hline
\multicolumn{5}{c}{\multirow{2}{*}{\textbf{LTR + SMOTE}}}  \\
\multicolumn{5}{c}{} \\
RankNet & 0.4762 / 0.5114 / 0.5603 & 0.4861 / 0.5937 / 0.6557 & 0.3584 / 0.3695 / 0.3978 & 0.6279 / 0.6371 / 0.7041 \\
RankMSE & 0.4286 / 0.4897 / 0.5351 & 0.5000 / 0.5877 / 0.6524 & 0.3640 / 0.3823 / 0.4085 & 0.6114 / 0.6263 / 0.6912  \\
ListNet & 0.4762 / 0.4959 / 0.5523 & 0.4861 / 0.6095 / 0.6603 & 0.3737 / 0.3789 / 0.4061 & 0.6207 / 0.6297 / 0.6928 \\
LambdaRank & 0.4622 / 0.5064 / 0.5469 & 0.5972 / 0.6452 / 0.6861 & 0.3479 / 0.3576 / 0.3871 & 0.6217 / 0.6359 / 0.700 \\ 
\hline
\multicolumn{5}{c}{\multirow{2}{*}{\textbf{LTR + GMVAE}}}  \\
\multicolumn{5}{c}{} \\
RankNet & 0.4930 / 0.5006 / 0.5412 & 0.5417 / 0.6352 / 0.6886 & 0.3570 / 0.3739 / 0.3989 & 0.6056 / 0.6281 / 0.6967 \\
RankMSE & 0.4454 / 0.4861 / 0.5168 & 0.4861 / 0.5617 / 0.6348 & 0.3582 / 0.3630 / 0.3879 & 0.6036 / 0.6286 / 0.6946 \\
ListNet & 0.4622 / 0.4820 / 0.5274 & 0.4167 / 0.5467 / 0.6327 & 0.3832 / 0.3829 / 0.4043 & 0.5997 / 0.6261 / 0.6943\\
LambdaRank & 0.4762 / 0.4873 / 0.5369 & 0.5556 / 0.6237 / 0.6844 & 0.3627 / 0.3827 / 0.4091 & 0.5927 / 0.6187 / 0.6885 \\ 
\hline
\multicolumn{5}{c}{\multirow{2}{*}{\textbf{LTR + Policy-Gradient}}}  \\
\multicolumn{5}{c}{} \\
PL-Rank-1 & 0.4416 / 0.5046 / 0.5342 & 0.5611 / 0.6133 / 0.6569 & 0.3489 / 0.3614 / 0.3906 & 0.6051 / 0.6110 / 0.6764\\
PL-Rank-2 & 0.4400 / 0.4975 / 0.5275 & 0.5485 / 0.6124 / 0.6502 & 0.3448 / 0.3591 / 0.3822 & 0.6129 / 0.6159 / 0.6788\\
PL-Rank-3 & 0.4458 / 0.5036 / 0.5426 & 0.5563 / 0.6209 / 0.6825 & 0.3455 / 0.3649 / 0.3958 & 0.6108 / 0.6284 / 0.6921 \\
\hline
\multicolumn{5}{c}{\multirow{2}{*}{\bf{LTR + Unbiased}}}  \\
\multicolumn{5}{c}{} \\
IPS & 0.3921 / 0.4866 / 0.5154 & 0.6102 / 0.6514 / 0.6968 & 0.3383 / 0.4008 / 0.4100 & 0.6497 / 0.6683 / 0.7019\\
DM & 0.3808 / 0.4715 / 0.5042 & 0.5994 / 0.6491 / 0.6979 & 0.3246 / 0.3918 / 0.4289 & \textbf{0.6597} / 0.6512 / 0.7048\\
DR & 0.4226 / 0.4736 / 0.5159 & 0.6108 / 0.6465 / 0.6985 & 0.3613 / 0.3981 / 0.4246 & 0.6535 / \textbf{0.6817} / 0.7002\\ 
\hline
\multicolumn{5}{c}{\multirow{2}{*}{\bf{MLTR + without Fine-tuning}}}  \\
\multicolumn{5}{c}{} \\
RankNet & $\ddagger$0.4874 / $\ddagger$0.4895 / $\ddagger$0.5444 & $\ddagger$0.5694 / $\ddagger$0.6048 / $\ddagger$0.6667 & $\ddagger$0.3592 / $\ddagger$0.3702 / $\ddagger$0.3991 & 0.6159 / 0.6339 / 0.7028 \\
RankMSE & $\ddagger$0.4454 / $\ddagger$0.4788 / $\ddagger$0.5191 & $\ddagger$0.5833 / $\ddagger$0.5964 / $\ddagger$0.6544 & $\ddagger$0.3867 / $\ddagger$0.3917 / $\ddagger$0.4158 & 0.6275 / 0.6400 / 0.7082 \\
ListNet & $\ddagger$0.5042 / $\ddagger$0.5068 / $\ddagger$0.5519 & $\ddagger$0.5694 / $\ddagger$0.6116 / $\ddagger$0.6669 & 0.4002 / 0.3996 / 0.4230 & $\ddagger$0.6336 / 0.6465 / $\ddagger$0.7141 \\
LambdaRank & $\ddagger$0.5014 / $\ddagger$0.5139 / $\ddagger$0.5669 & $\ddagger$0.6250 / $\ddagger$0.6504 / $\ddagger$0.6981 & 0.3662 / 0.3798 / 0.4092 & 0.6072 / 0.6307 /	0.7024 \\
\hline
\multicolumn{5}{c}{\multirow{2}{*}{\bf{MLTR + with Fine-tuning}}}  \\
\multicolumn{5}{c}{} \\
RankNet & $\ddagger$\textbf{0.5770} / $\ddagger$\textbf{0.5460} / $\ddagger$0.5913 & $\ddagger$0.6250 / $\ddagger$0.6452 / $\ddagger$0.6949 & $\ddagger$0.3873 / $\ddagger$0.3889 / $\ddagger$0.4122 & $\ddagger$0.6212 / $\ddagger$0.6385 / $\ddagger$0.7071 \\
RankMSE & $\ddagger$0.4902 / $\ddagger$0.5238 / $\ddagger$0.5646 & $\ddagger$0.5972 / $\ddagger$0.6505 / $\ddagger$0.6985 & $\ddagger$0.3887 / $\ddagger$0.3981 / $\ddagger$0.4237 & $\ddagger$0.6362 / $\ddagger$0.6450 / $\ddagger$0.7112 \\
ListNet & $\ddagger$0.5266 / $\ddagger$0.5254 / $\ddagger$0.5704 & $\ddagger$0.6111 / $\ddagger$0.6473 / $\ddagger$0.7027 & $\ddagger$\textbf{0.4088} / \textbf{$\ddagger$0.4100} / $\ddagger$\textbf{0.4342} & $\ddagger$0.6388 / $\ddagger$0.6490 / $\ddagger$\textbf{0.7172} \\
LambdaRank & $\ddagger$0.5350 / $\ddagger$0.5409 / $\ddagger$\textbf{0.5914} & $\ddagger$\textbf{0.6389} / $\ddagger$\textbf{0.6590} / $\ddagger$\textbf{0.7130} & $\ddagger$0.3739 / $\ddagger$0.3850 / $\ddagger$0.4119 & $\ddagger$0.6196 / $\ddagger$0.6397 /	$\ddagger$0.7089\\
\bottomrule
   \end{tabular}
}
\label{experimental_results}
\end{table}

\label{sec:experi_baseline}
We compare the performance of MLTR to traditional LTR models when handling sparsely labeled queries. We tested our models on the four public datasets by simulating sparse data scenarios. The process involved training on $\mathcal{S}_{train: p1n9}$ and $\mathcal{S}_{test: p1n9}$, followed by fine-tuning on $\mathcal{T}_{tuning: p1n9}$ and evaluating the results on $\mathcal{T}_{eval: rest}$. The results shown in Table~\ref{experimental_results} indicate that our MLTR models, regardless of being in a without fine-tuning or with fine-tuning setting, outperform the baseline models in all metrics (NDCG@1, NDCG@5, and NDCG@10) across all four datasets, with the exception of the unbiased click-based LTR baseline models. This improved performance is maintained across various loss functions. Notably, on the MQ2007 and MQ2008 datasets, where the positive sample distribution is relatively sparse, the MLTR model shows significant improvement across all loss functions, providing further evidence that the meta-learning approach can enhance the model's predictive ability in sparse data. While the MSLR dataset has a relatively even distribution of positive and negative samples overall, the MLTR model still manages to improve the model's predictive results in most of the loss functions. The results also suggest that the SMOTE resampling technique can alleviate the issue of sparse data and improve performance compared to the traditional LTR model. The GMVAE-based data augmentation method outperforms the SMOTE resampling method in most cases as it can incorporate more informative data. However, data augmentation-based methods do not significantly enhance model generalization compared to our proposed MLTR approach. In comparing the PL-rank methods with MLTR, we found that the PL ranking methods lacks consistent stability, particularly underlined by our experiments that highlight the sparse nature of positive samples during training. In contrast, the test results indicate that MLTR yields more stable outcomes in scenarios characterized by a limited number of training samples or a sparser distribution of positive samples. On the other hand, when examining the results of Unbiased click-based LTR methods, these methods show a notable advantage when the training samples contain rich features in the query and document pairs. For instance, on the Istella-S dataset, DM and DR achieved the best performance in NDCG@1 and NDCG@5. However, in the other three datasets, the MLTR model displayed a more consistent performance advantage. In our proposed MLTR model, we did not strictly address the bias present in the data. The experimental outcomes of unbiased click-based LTR methods indicate that incorporating unbiased methods like IPS into the training process of meta-learning might further enhance the performance of meta-learning-based LTR models. We plan to implement and evaluate this approach in our future work. On the other hand, regardless of whether the MLTR model utilizes the fine-tuning process during meta-testing, it consistently demonstrates competitiveness compared to traditional methods. The results of MLTR without fine tuning still lead in most experiments, surpassing traditional LTR models as well as baseline models with other optimization approaches. Furthermore, if the fine-tuning process in meta-testing is employed, we find that MLTR can adapt more rapidly to changes in queries, thereby further enhancing the model's performance on test evaluation data. When dealing with sparsely labeled queries, our MLTR model can achieve better adaptability with a small proportion of labeled data, leading to improved overall model performance.

\begin{figure}[tb]
     \centering
     \begin{subfigure}[b]{0.8\textwidth}
         \centering
         \includegraphics[width=\textwidth]{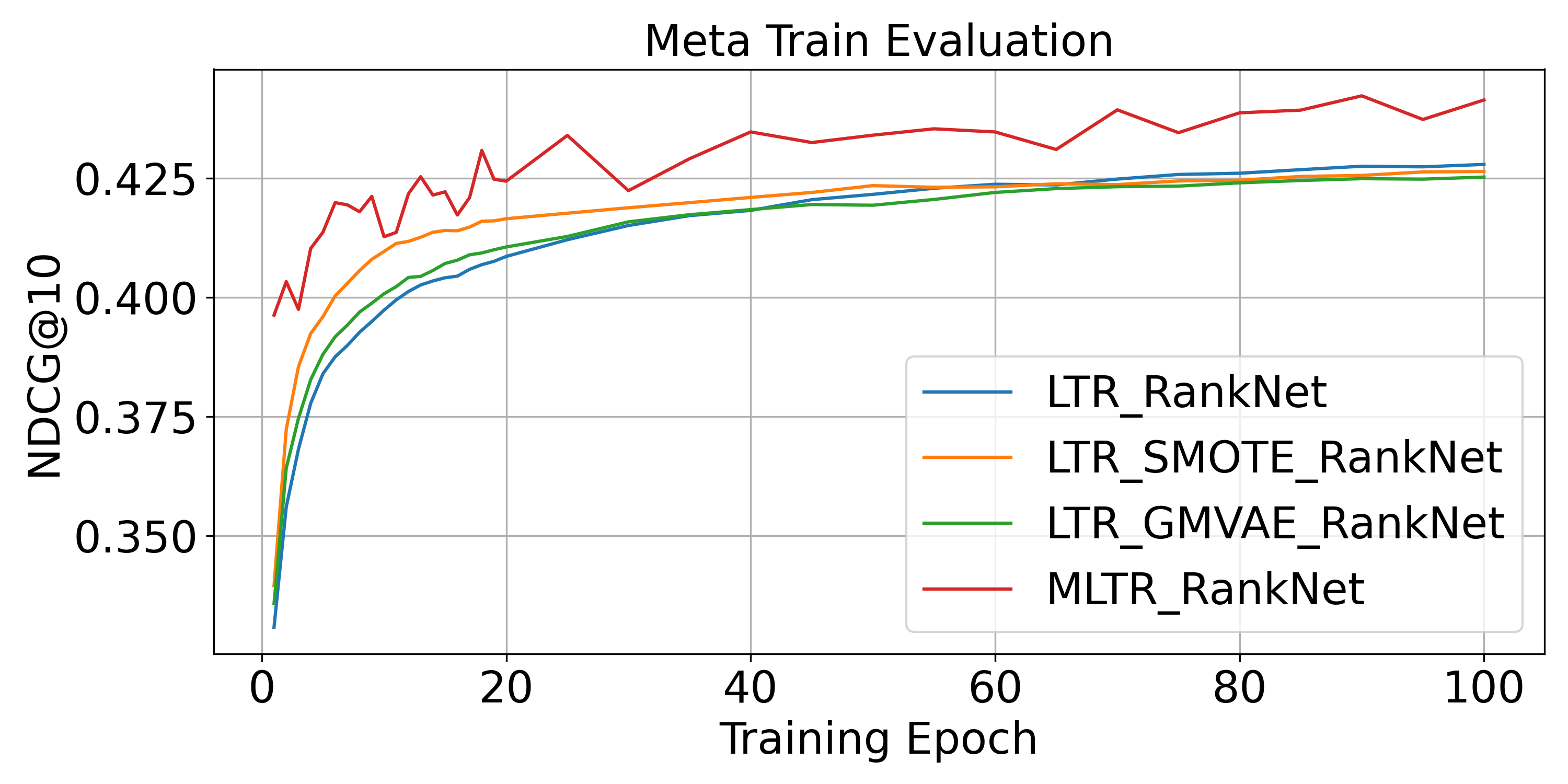}
         \caption{Meta train evaluation based on meta test single fine-tuning}
         \label{fig:train_eval}
     \end{subfigure}
     % \hfill
     \begin{subfigure}[b]{0.8\textwidth}
         \centering
         \includegraphics[width=\textwidth]{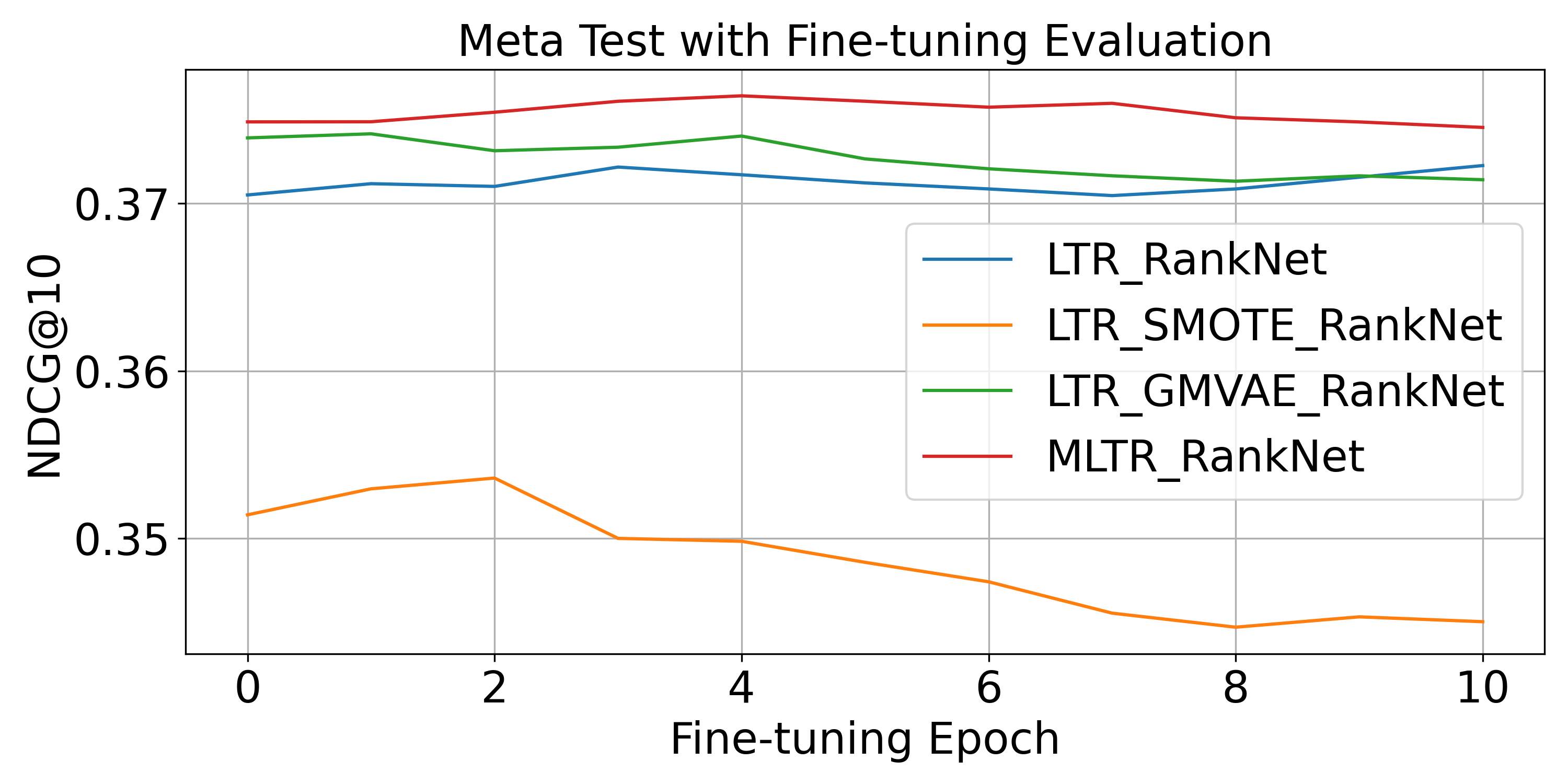}
         \caption{Meta test fine-tuning evaluation with the best training model}
         \label{fig:test_eval}
     \end{subfigure}
     % \hfill
        \caption{Meta train/test evaluation on NDCG@10 of MLTR and other baselines with RankNet on the MSLR-10K dataset}
        \label{fig:train_test_eval}
\end{figure}

\begin{figure}  %[htbp]
     \centering
     \begin{subfigure}[b]{\textwidth}
         \centering
         \includegraphics[width=\textwidth]{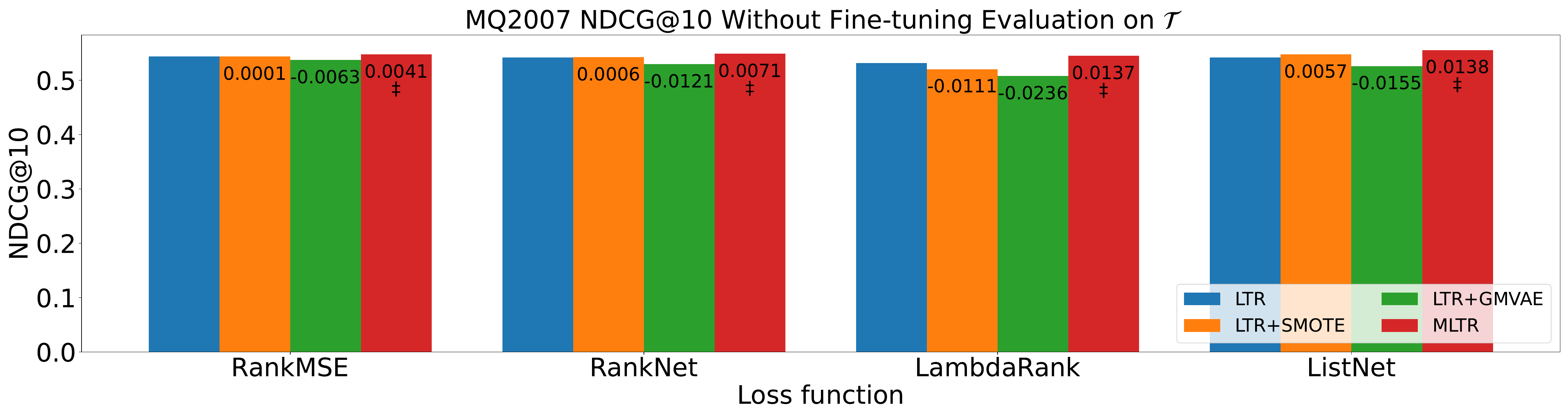}
         \caption{MQ2007}
         \label{fig:finetuning_MQ2007}
     \end{subfigure}
     % \hfill
     \begin{subfigure}[b]{\textwidth}
         \centering
         \includegraphics[width=\textwidth]{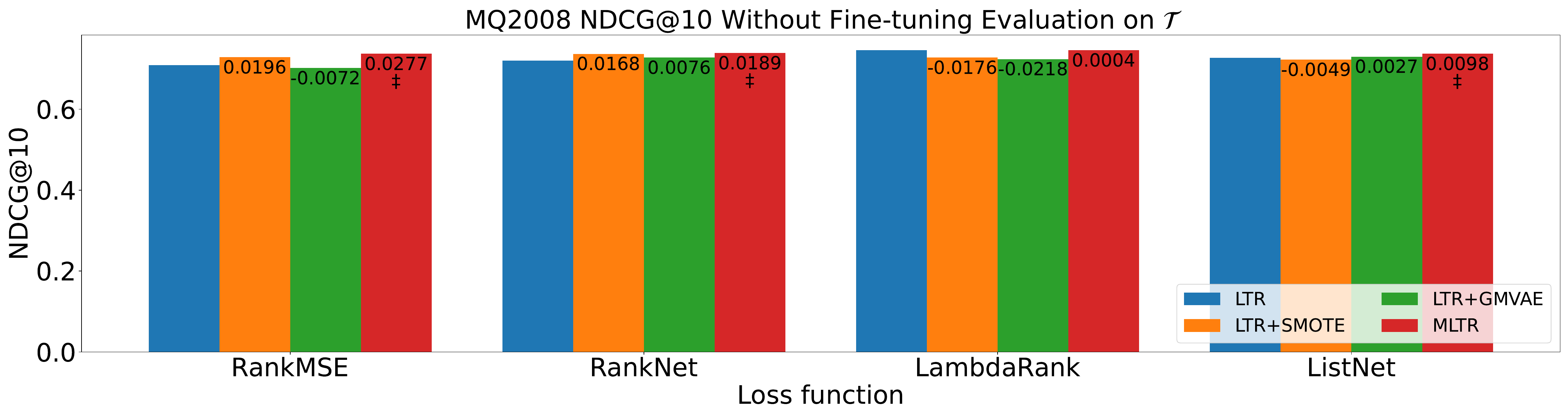}
         \caption{MQ2008}
         \label{fig:finetuning_MQ2008}
     \end{subfigure}
     % \hfill
     \begin{subfigure}[b]{\textwidth}
         \centering
         \includegraphics[width=\textwidth]{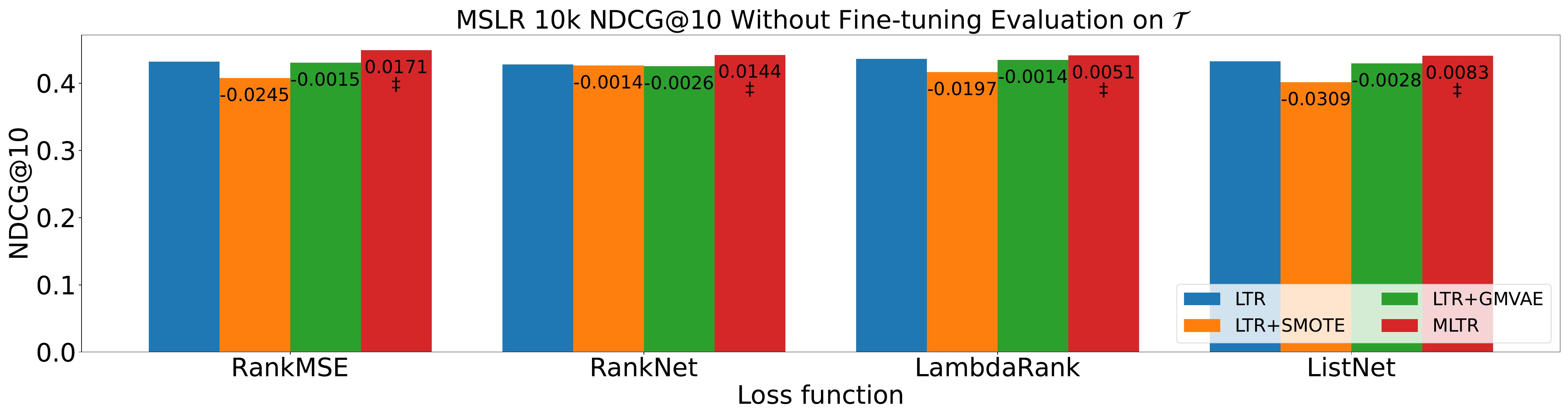}
         \caption{MSLR-10K}
         \label{fig:finetuning_mslr}
     \end{subfigure}
     \begin{subfigure}[b]{\textwidth}
         \centering
         \includegraphics[width=\textwidth]{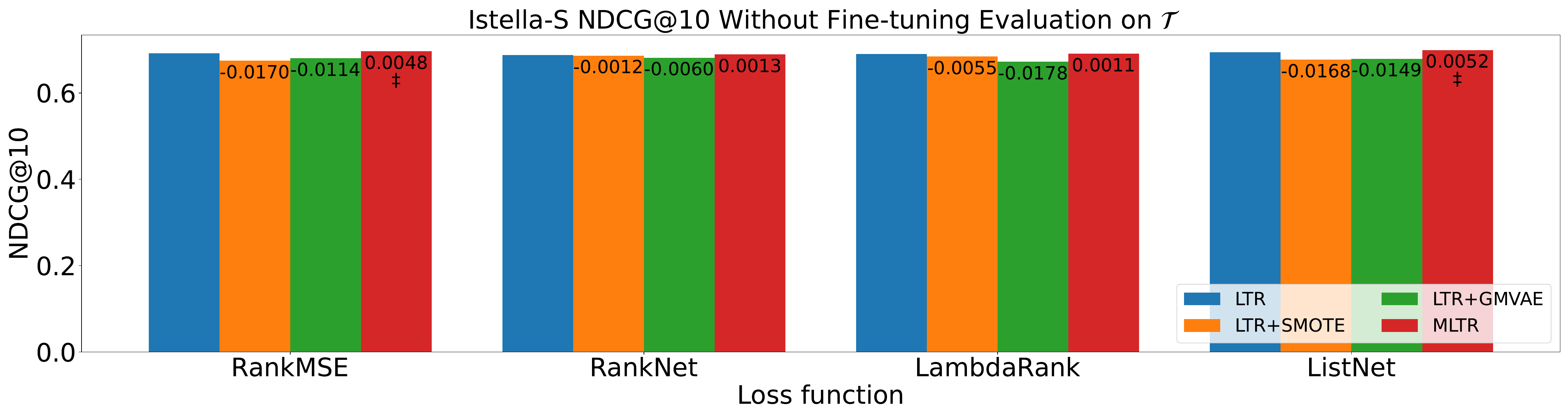}
         \caption{Istella-S}
         \label{fig:finetuning_istella}
     \end{subfigure}
        \caption{Comparison the performance of models without fine-tuning, using various loss functions and models, on the NDCG@10 metric of the entire test dataset $\mathcal{T}$ from four different public datasets. The symbol $\ddagger$ in the bar indicates a statistically significant improvement of MLTR without fine-tuning over the corresponding LTR models, as evidenced by a p-value $<$ 0.01 in a two-tailed t-test.}
        \label{fig:public_sample}
\end{figure}

\subsection{Ablation Study}

\label{sec:ablation}

\subsubsection{The Effect of Meta Train and Meta Test (RQ2)}
\label{sec:meta_train_test}

Fig. \ref{fig:train_test_eval} illustrates the performance trend of the NDCG@10 metric on the test data during the training process (as shown in Fig. \ref{fig:train_eval}) and the fine-tuning process (as shown in Fig. \ref{fig:test_eval}) for both MLTR and baseline models based on the RankNet loss. In Fig. \ref{fig:train_eval}, NDCG@10 of the test data is calculated by fine-tuning the model for one epoch on the meta-test tuning data after each training epoch ($1 \leq e \leq 100$) during the training process. The results show that MLTR consistently outperforms the LTR and other baseline models during the training process and can achieve close to its best performance within only a few epochs.

Fig.~\ref{fig:test_eval} demonstrates the performance of MLTR and baseline models on the test data during the fine-tuning process. The model (the best model from meta training stage) was fine-tuned for 10 epochs on the meta-test test data, with NDCG@10 computed for each epoch. The results show that MLTR consistently performs better than the LTR and other baseline models throughout the fine-tuning process. Our model demonstrates clear and stable performance on unseen datasets through a straightforward fine-tuning process, mitigating the effects of label imbalance and potential domain shifts. Additionally, MLTR still significantly outperforms the baseline models even without any fine-tuning on the meta-test test data, as shown by the NDCG@10 metrics at epoch 0 in Fig.~\ref{fig:test_eval}. During fine-tuning, MLTR continues to improve and outperform the corresponding baseline models under the RankNet loss from epoch 0 to epoch 4. On the other hand, the baseline models tend to suffer from overfitting problems, resulting in a decline in performance.

\begin{table}[t]
\caption{Comparative Analysis of LTR and MLTR Frameworks Using NDCG@1, NDCG@5, and NDCG@10 Metrics on the MSLR-10K Dataset. This analysis reflects a training approach where each query is paired with one positive document and a random number of negative documents. The evaluation is conducted consistently on the same dataset.}
\resizebox{\textwidth}{!}{%
\begin{tabular}{llcccc}
\toprule
\multicolumn{1}{c}{\multirow{2}{*}{\textbf{Model}}} & \multicolumn{1}{c}{\multirow{2}{*}{\textbf{Method}}} & \multicolumn{2}{c}{\textbf{Without Fine Tuning}}                                       & \multicolumn{2}{c}{\textbf{With Fine Tuning}}                                          \\
\multicolumn{1}{c}{}                       & \multicolumn{1}{c}{}                        & \textbf{NDCG@1 / 5 / 10} & \textbf{Percentage Increase} & \textbf{NDCG@1 / 5 / 10} & \textbf{Percentage Increase} \\
\hline
\textbf{LTR}                               & \textbf{LambdaRank}                         & 0.3359 / 0.3757 / 0.4086 &      -   /    -     /  -           & 0.3411 / 0.3809 / 0.4161    &        - /    -     /   -          \\
\textbf{MLTR}                              & \textbf{LambdaRank}                         & 0.3555 / 0.3816 / 0.4158    & 5.83\%  / 1.58\%  / 1.76\%      & 0.3582 / 0.3822 / 0.4181    & 5.00\%  / 0.34\%  / 0.48\%      \\
\hline
\textbf{LTR}                               & \textbf{ListNet}                            & 0.3290 / 0.3703 / 0.4087    &    -     /    -     /  -           & 0.3328 / 0.3729 / 0.4107    &     -    /     -    /    -         \\
\textbf{MLTR}                              & \textbf{ListNet}                            & 0.3632 / 0.3898 / 0.4259    & 10.37\% / 5.26\%  / 4.22\%      & 0.3634 / 0.3898 / 0.4268    & 9.20\%  / 4.55\%  / 3.91\%      \\
\hline
\textbf{LTR}                               & \textbf{RankMSE}                            & 0.3129 / 0.3418 / 0.3777    &     -    / -        /   -          & 0.3136 / 0.3430 / 0.3785    &      -   /     -    /      -       \\
\textbf{MLTR}                              & \textbf{RankMSE}                            & 0.3631 / 0.3868 / 0.4183    & 16.06\% / 13.17\% / 10.76\%     & 0.3631 / 0.3868 / 0.4188    & 15.78\% / 12.78\% / 10.66\%     \\
\hline
\textbf{LTR}                               & \textbf{RankNet}                            & 0.3188 / 0.3545 / 0.3919    &     -    /   -      /  -           & 0.3191 / 0.3547 / 0.3923    &      -   /    -     /    -         \\
\textbf{MLTR}                              & \textbf{RankNet}                            & 0.3462 / 0.3805 / 0.4107    & 8.57\%  / 7.32\%  / 4.79\%      & 0.3463 / 0.3805 / 0.4108    & 8.50\%  / 7.26\%  / 4.71\% \\
\bottomrule
\end{tabular}
}
\label{tb:mltr_with_random_neg}
\end{table}

\subsubsection{Meta Test without Fine-tuning (RQ3)}
\label{sec:without_fine_tuning}

In this section, we delve deeper into the results of our model's ability to maintain competitiveness on unseen datasets without fine-tuning. Fig. \ref{fig:public_sample} provides a comparison of the performance of our MLTR model with the baseline models on various datasets using the same entire test dataset $\mathcal{T} = \mathcal{T}_{tuning} \cup \mathcal{T}_{eval}$. As we can see, our model still has stronger prediction ability for unseen data or distribution compared to the other models. We have calculated the absolute growth of the data-augmentation based model and MLTR as compared to the baseline LTR across multiple metrics. When comparing with other models, we can see that MLTR consistently outperforms the LTR model across all 16 experiments, which encompass 4 datasets and 4 different loss functions. It is evident that the red bars, symbolizing MLTR, consistently exhibit an increase in performance in all comparative experiments relative to other methods. While the absolute magnitude of this growth may not appear substantial, the consistent improvement observed across four distinct datasets and four diverse optimization ranking functions underscores the robust reliability of the MLTR approach. Additionally, the significant test results further demonstrate that the majority of these improvements are statistically significant. This consistency aligns with the results shown in Table \ref{experimental_results}. On the other hand, we noted that the data augmentation-based LTR models, which aimed to rebalance the ratio of positive and negative samples in the training set using synthetic data, did not uniformly improve performance during testing. In fact, on some metrics, their performance was even worse than traditional LTR models. Furthermore, the results from these datasets highlight that while data augmentation helps mitigate the impact of imbalanced positive and negative samples, it does not effectively enhance the model's generalization ability when facing domain shift issues in testing.

\subsubsection{MLTR with Query-Document Pairs (RQ3)}
\label{sec:MLTR_qd_pair}

To better validate the universality of the MLTR model, we introduced a new set of comparative experiments within the MLTR-10K dataset. For each query, we randomly selected 2 positive samples and 78 negative samples. During the meta-training process, instead of adhering to a fixed p1n39 positive-negative ratio, we opted for a variable number of negative samples while keeping one positive sample constant. This experimental setting is designed to test whether MLTR outperforms LTR in scenarios with varying numbers of documents per query. The results in Table \ref{tb:mltr_with_random_neg} demonstrate that MLTR, even with dynamically adjusted numbers of documents per query, still shows a significant advantage over LTR methods across different optimization methods. We also compared results before and after fine-tuning. These results further confirm that MLTR consistently outperforms LTR models, regardless of fine-tuning, underscoring MLTR's superior adaptability to new tasks.

\subsection{Robustness of MLTR (RQ4)}
\label{sec:robustness}

This section demonstrates the superiority of our meta-learning model over the baseline when dealing with extremely sparse data and a low positive-to-negative label ratio. The evaluation was conducted on the MSLR-10K dataset and various experimental scenarios were simulated by sampling subsets of the data with varying ratios of positive and negative labels per query.

\subsubsection{Experiment Setup}

\begin{figure}[t]  %[htbp]
     \centering
     \includegraphics[width=0.9\textwidth]{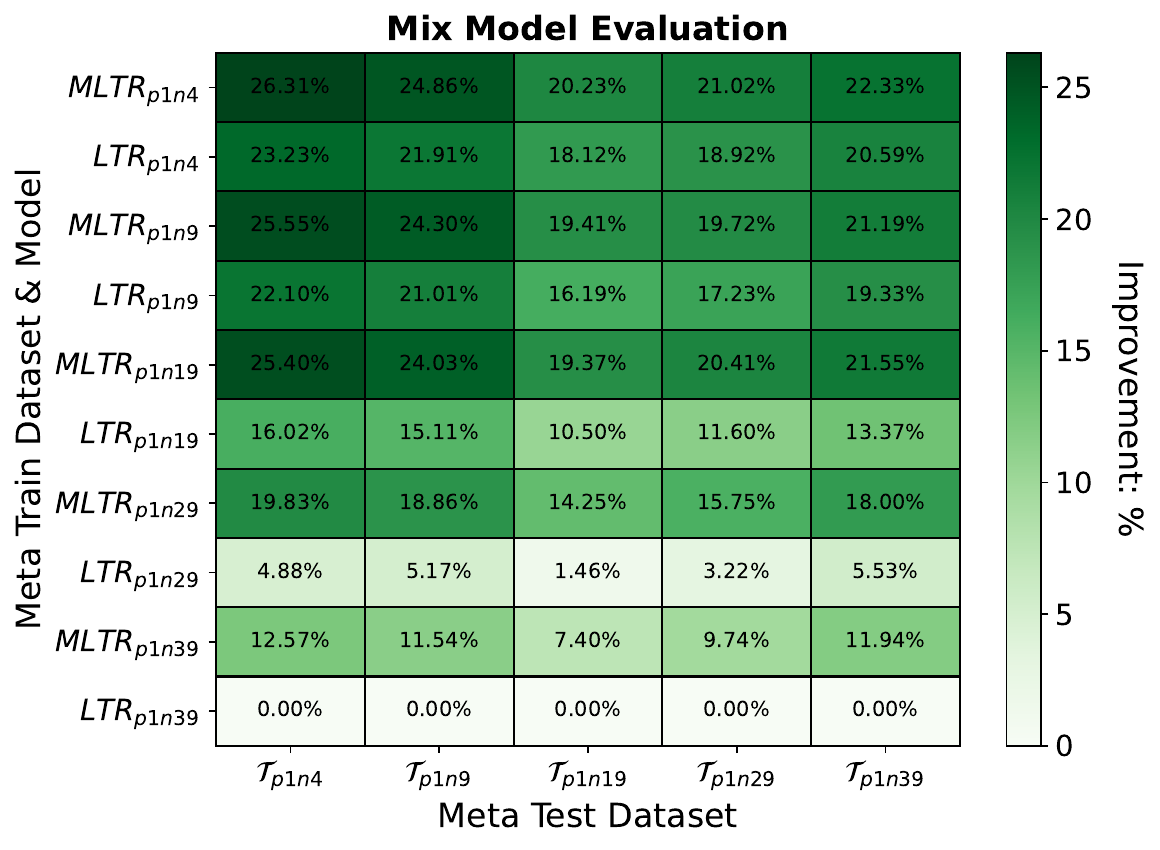}
     \caption{Relative improvement experimental results of NDCG@10 from MLTR and LTR based on RankNet loss in variant sparsely labeled data setting on MSLR-10K dataset}
     \label{fig:pub_mix_model_eval}
\end{figure}

The factors to consider in this experiment include the number of sampled positive/negative labeled items per query in the training data, 
the number of sampled positive/negative labeled items per query in the test data, and the training model.
For the model training, we compare the model performance between
the baseline LTR model and the MLTR model with RankNet loss for both models,
denoted by MLTR and LTR respectively. We use NDCG@10 as the evaluation metric. There is no overlap between any sampled training and test data in order to ensure the fairness of the experiment. The sampled training and test data are introduced in Section~\ref{sample_data}.
Given a combination of the training model, $\mathcal{S}$, $\mathcal{T}$, 
we can obtain the NDCG@10 metrics on $\mathcal{T}_{eval}$ with the best model trained on $\mathcal{S}$ and fine-tuned on $\mathcal{T}_{tuning}$ with 1 epochs.
The experimental results for all combinations are shown in Fig. \ref{fig:pub_mix_model_eval}. 

\subsubsection{Data Distribution Shift Evaluation}
\label{sec:relative_improvement}

Fig.~\ref{fig:pub_mix_model_eval} shows the relative NDCG@10 gain on the worst-performing model $LTR_{p1n39}$.
For each test data $\mathcal{T}$ (represented by x-axis) corresponding to a column, 
the NDCG@10 metrics are computed for a model (LTR or MLTR) trained on $\mathcal{S}$ (represented by y-axis). $LTR_{p \cdot n \cdot}$ and $MLTR_{p\cdot n \cdot}$ denote the model LTR and MLTR trained on specific training datasets, respectively;
each grid in this column shows the relative NDCG@10 improvement compared against 
$LTR_{p1n39}$ on the evaluation data $\mathcal{T}_{eval}$. 
The darker the color of each grid, the greater the improvement of the model in this grid relative to $LTR_{p1n39}$.

We have the following observations on Fig.~\ref{fig:pub_mix_model_eval}.
First, for any given $\mathcal{S}$ and $\mathcal{T}$, MLTR performs better than LTR consistently, with a significant 1.86\% - 14.95\% improvement. 
For example, $MLTR_{p1n29}$ improves 12.53\% over $LTR_{p1n29}$ on the test data $\mathcal{T}_{p1n29}$ (15.75\% for MLTR vs. 3.22\% for LTR shown in Fig.~\ref{fig:pub_mix_model_eval}).
Second, we observe that MLTR is much more stable and robust to the sparse data than LTR, 
by comparing all models' performance for a fixed test data $\mathcal{T}$ (corresponding to a column). 
As $\mathcal{S}$ gets more sparse, performance degradation is observed for both MLTR and LTR models; 
however, MLTR's NDCG performance decreases much slower compared to LTR, 
For example, looking at these models' NDCG metrics on $\mathcal{T}_{p1n4}$ (corresponding to column 1) 
as the training data get more sparse from $\mathcal{S}_{p1n4}$ to $\mathcal{S}_{p1n39}$, 
NDCG for LTR decreases by 23.23\%, from 23.23\%  ($LTR_{p1n4}$) to 0\% ($LTR_{p1n39}$), 
while NDCG for MLTR decreases by 13.74\% from 26.31\% ($MLTR_{p1n4}$) to 12.57\% ($MLTR_{p1n39}$).
In addition, as pointed out in Section~\ref{sec:robustness}, all the models evaluated in Fig.~\ref{fig:pub_mix_model_eval}
go through only one training epoch on $\mathcal{T}_{tuning}$. 
With MLTR's significant improvement over LTR under all the scenarios, we show that the meta-based LTR models
can generalize and adapt significantly better than the traditional LTR models under sparsely labeled data settings.

\subsection{Real-world Application Case Study (RQ5)}
\label{sec:case_study}
The study is performed on the real-world Walmart.com dataset, which has sparse positive-labeled queries as shown in Table \ref{tb:staticstic}. It is worth conducting the robustness experiments to evaluate the model generalization, as the data in the real world are often more dynamic with drifted distributions. 

\begin{table}[t]
\caption{NDCG@1 and NDCG@5 Gain are reported in terms of the percentage lift for MLTR over LTR on various loss of Walmart dataset, $\ddagger$ denotes statistically significant improvement from LTR to MLTR with the p-value < 0.01 using the two-tailed t-test}
\begin{tabular}{ccc}
\toprule
\bf Loss & \bf Gain NDCG@1 & \bf Gain NDCG@5 \\
\hline
RankMSE & $\ddagger$+0.44\% & $\ddagger$+0.95\% \\
RankNet & $\ddagger$+1.99\% & $\ddagger$+0.92\% \\
LambdaRank & $\ddagger$+2.58\% & $\ddagger$+1.04\% \\
ListNet & $\ddagger$+2.32\% & $\ddagger$+1.51\% \\
\bottomrule
\end{tabular}
\label{ndcg_gain_walmart}
\end{table}

\subsubsection{Experimental Results}

\begin{figure}
\centering
    \begin{subfigure}[b]{0.8\textwidth}
    \includegraphics[width=\textwidth]{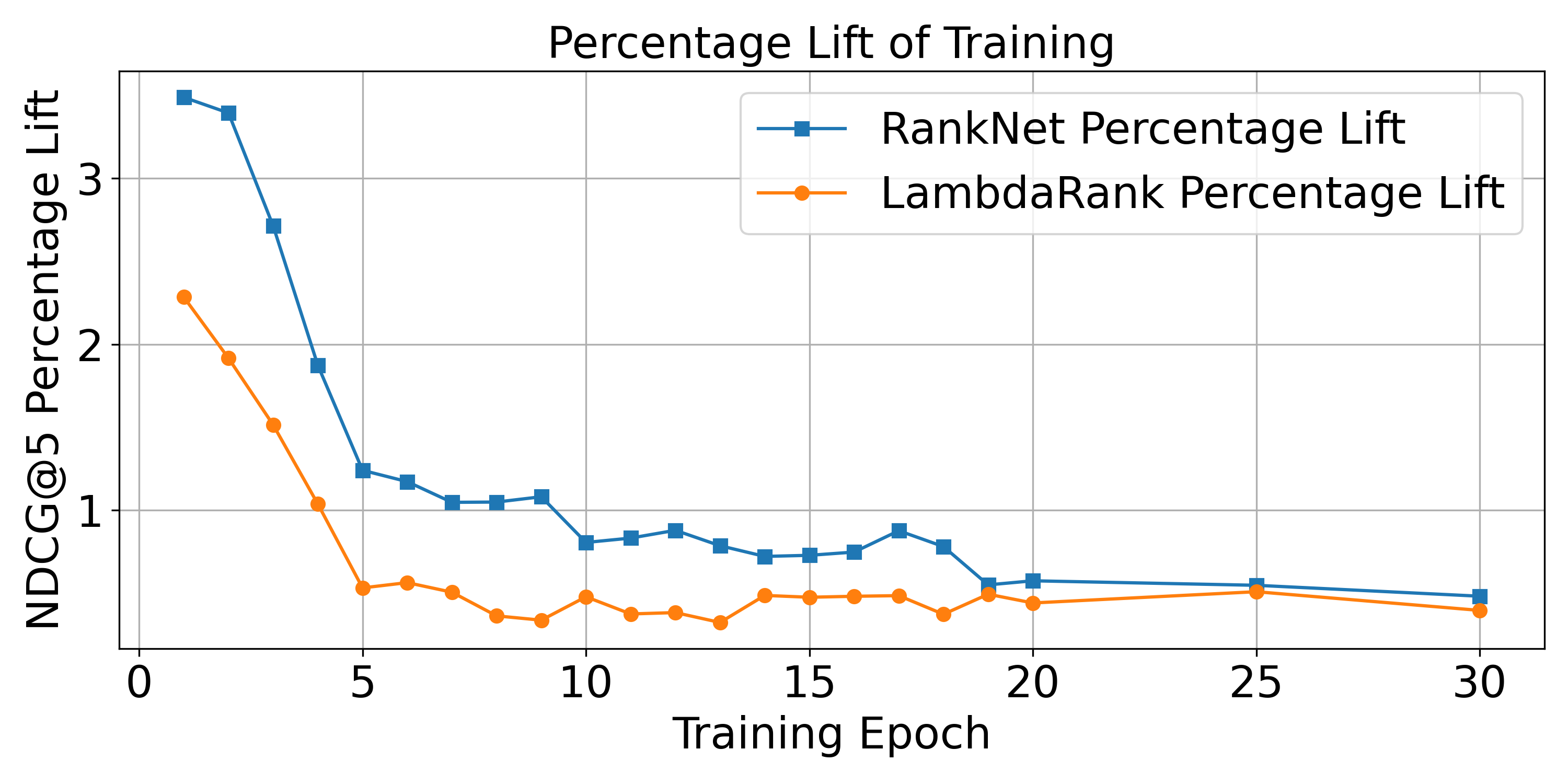}
    \caption{NDCG@5 percentage lift comparison between LTR and MLTR models based on 30 training epochs in meta train evaluation}
    \label{fig:walmart_train_eval}
    \end{subfigure}
    \begin{subfigure}[b]{0.8\textwidth}
    \includegraphics[width=\textwidth]{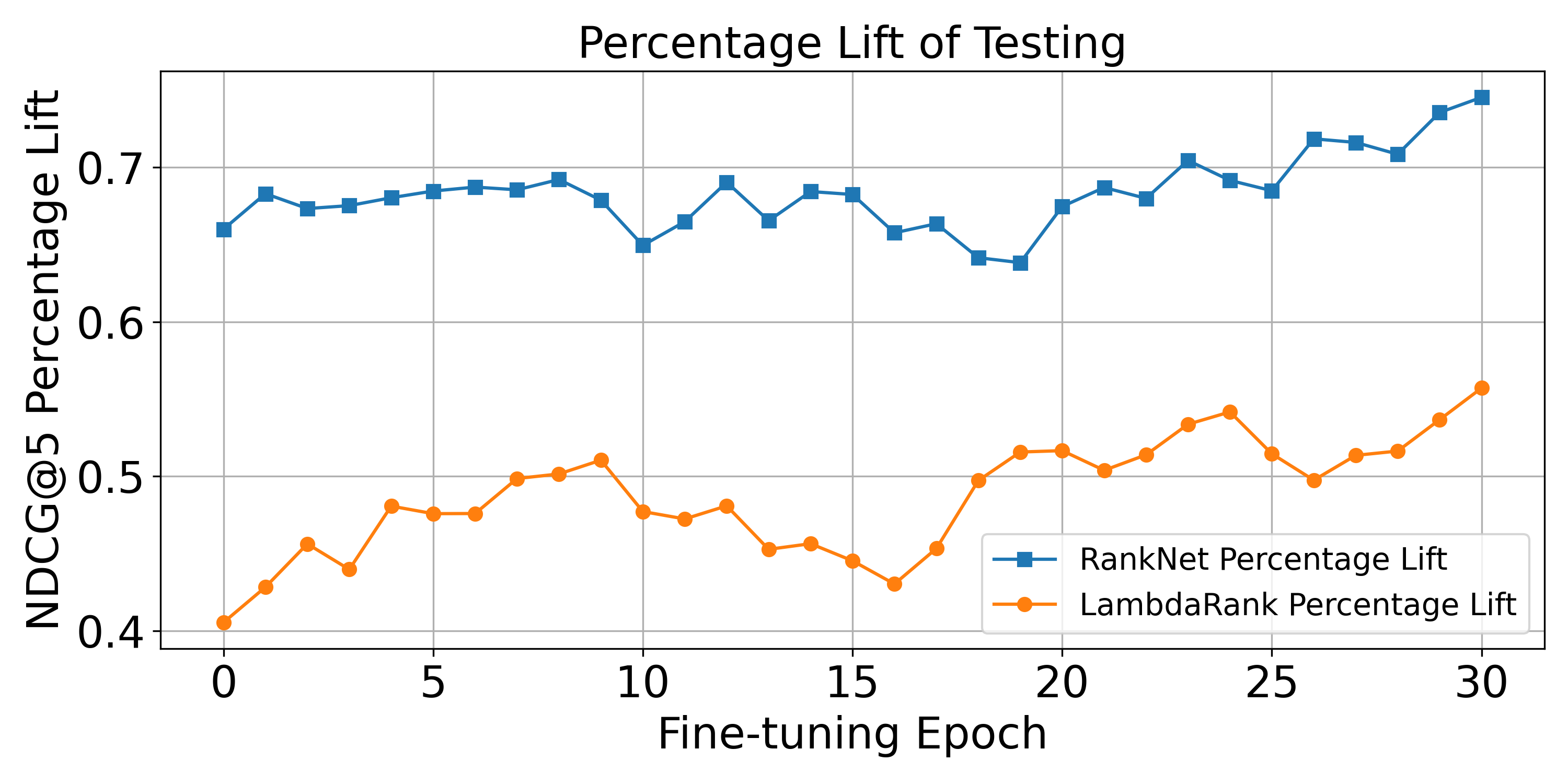}
    \caption{NDCG@5 percentage lift comparison between LTR and MLTR models based on 30 fine-tuning epochs in meta test evaluation}
    \label{fig:walmart_test_eval}
    \end{subfigure}
\caption{Meta train/test evaluation of NDCG@5 percentage lift for MLTR and LTR models using RankNet and LambdaRank on the Walmart.com dataset}
\label{fig:walmart_train_test_eval}
\end{figure}

Table \ref{ndcg_gain_walmart} shows the percentage lift in NDCG@1 and NDCG@5 for MLTR over LTR on the Walmart dataset for various loss functions. The results from the Walmart dataset align with the patterns observed in the public datasets. Our MLTR models outperform the traditional LTR models in terms of both NDCG@1 and NDCG@5 metrics in sparsely labeled data scenarios. On the other hand, Fig. \ref{fig:walmart_train_eval} illustrates the NDCG@5 gain between the MLTR and LTR models using RankNet and LambdaRank losses. The same training method as used on the public dataset was employed, with the model fine-tuned for one epoch on the test support data at the end of each training epoch ($1 \leq e \leq 30$) during the training process. The NDCG@5 gain was then computed based on $(MLTR_{NDCG@5}-LTR_{NDCG@5})/LTR_{NDCG@5}$. The results demonstrate that the MLTR consistently outperforms the LTR models in real-world application datasets. In the early stages of training, the MLTR model exhibits a greater improvement over LTR, demonstrating the efficiency of the MLTR model and its faster convergence speed. As the number of training iterations increases, both MLTR and LTR models become relatively stable, but the MLTR model still performs better than the traditional LTR model. This conclusion holds true for both the human-annotated relevance label sparsity setting seen in the three public datasets and the engagement label sparsity setting demonstrated in the Walmart dataset. During the testing process illustrated in Fig.~\ref{fig:walmart_test_eval}, the MLTR model consistently outperforms the LTR models throughout. Similar results were observed in the implementation on public datasets.

After comparing the NDCG@5 gain ratios during both the training and fine-tuning processes depicted in Fig.~\ref{fig:walmart_train_test_eval}, it can be observed that the MLTR model consistently outperforms the LTR models. During training, the MLTR model exhibits a significant improvement over LTR, between 0.33\% with 3.49\%. Similarly, during fine-tuning under similar conditions, the MLTR model achieves varying levels of performance improvement, ranging between 0.41\% with 0.73\%. Although the improvement ratio during training is not as pronounced, it still indirectly validates the efficiency and compatibility of the MLTR model with respect to the data and task.

\subsubsection{Sampling Strategy in Meta Training}

With the design of the inner loop during local updating on the meta-train data in the MLTR, 
we can sample a data subset for the model's local update with different sample strategies.
This sampling strategy during the local update aims to improve the model performance 
on the specific query-based tasks, and thus improves the model's generalization performance
on a new query.
In this section, in order to further boost the model performance under the data sparsity setting,
we explore different sampling strategies in the MLTR model by using subsets of the data as training data.
Several sampling strategies we investigate are defined as follows:

\begin{itemize}
\item \textbf{All data}: Use all data (256 items for each query) with baseline LTR model.
\item \textbf{Fixed Sampler}: Fixed sample 1 positive and 19 negatives data with MLTR model.
\item \textbf{1 Positive Sampler}: Randomly sample 1 positive and 19 negatives data each time in training with MLTR model.
\item \textbf{Multiple Positive Sampler}: Randomly sample 2 positives and 18 negatives data each time in training with MLTR model.
\end{itemize}

\begin{figure}
\centering
\includegraphics[width=0.8\columnwidth]{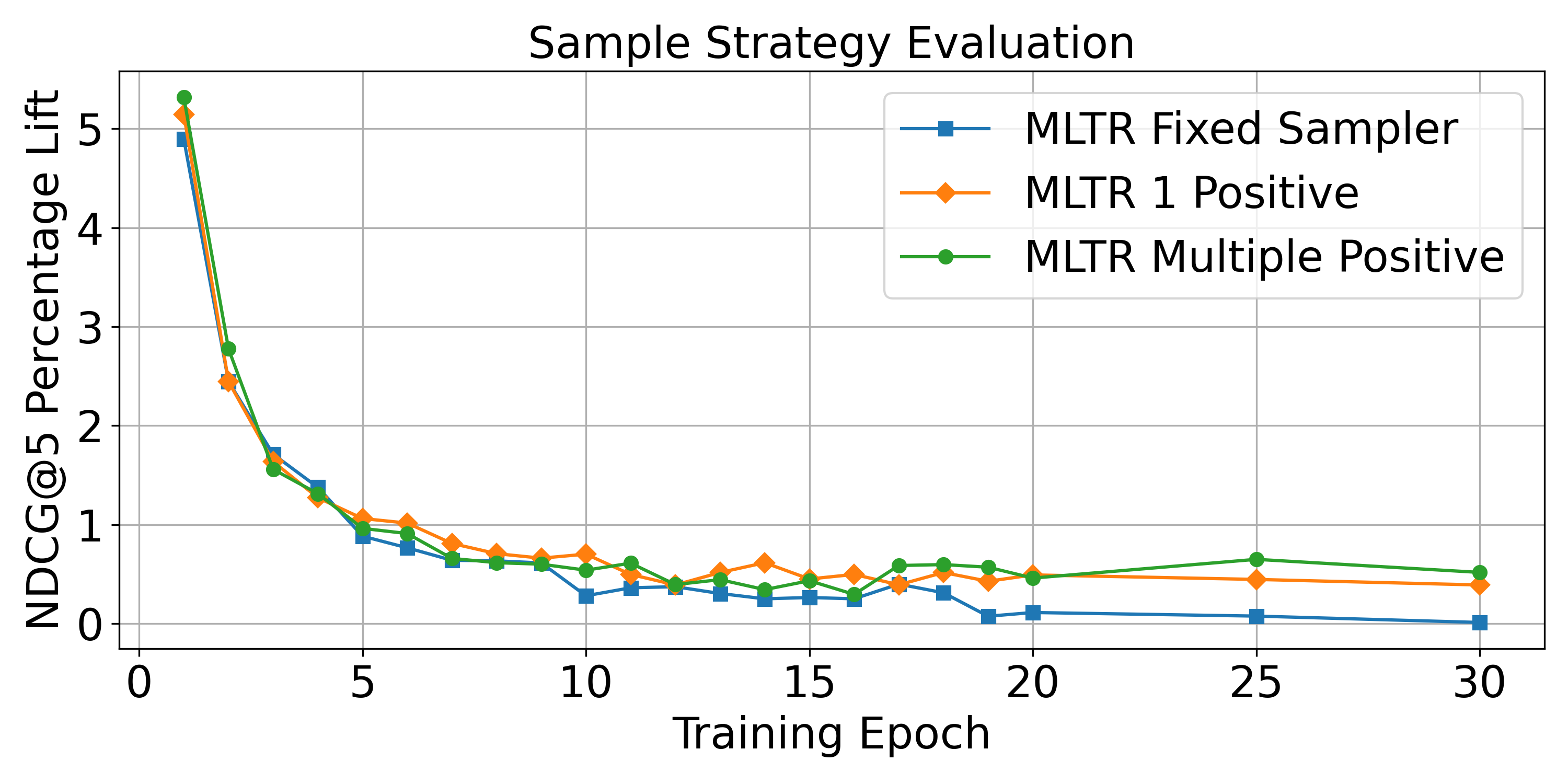}
\caption{NDCG@5 percentage lift in model performance using various sample strategies compared to $LTR_{All data}$ during the meta train evaluation of the MLTR model with RankNet on Walmart.com dataset}
\label{fig:sample}
\end{figure}

As shown in Fig.~\ref{fig:sample}, As the number of sampled positive data increases, the performance of the MLTR model shows a slight improvement, with the green line demonstrating the highest improvement, followed by the yellow line, and then the blue line. In addition, all the MLTR models with sampling strategies (green, yellow, and blue lines)
outperform the baseline LTR model with \textbf{All data} (the NDCG@5 percentage lift above 0). 
We can see that compared to using all the data for training LTR models, 
using variant partial of the data for each inner loop during the meta training 
can not only reduce the amount of computation during the training, 
but also improve the model generalization.

\subsubsection{BERT with Fine-tuning}

As we mentioned in the previous section, we extend the features with query and document text embedding representation;
this semantics information is available in the e-commerce dataset as the query and item title. 
We generate query text embedding and item title text embedding from the pre-trained distilled BERT model distillbert-base-uncased\footnote{\url{https://huggingface.co/distilbert-base-uncased}}; then the BERT model parameters are fine-tuned during the meta-learner update, similar as MeLu  \citep{DBLP:conf/kdd/LeeIJCC19}.
No significant improvement is observed by using the Bert-based query/item text embeddings. 
This is not surprising since the BERT-based embedding information for a \(query, production\) pair
is already covered by a numeric feature in the e-commerce dataset, which is computed as the
cosine similarity between a Bert-based query embedding vector and item embedding vector \citep{DBLP:conf/kdd/MagnaniLCYSPC0K22}.

\subsubsection{Data Distribution Shift in Walmart.com}

\begin{figure}[t]  %[htbp]
     \centering
     \begin{subfigure}[b]{0.9\textwidth}
         \centering
         \includegraphics[width=\linewidth]{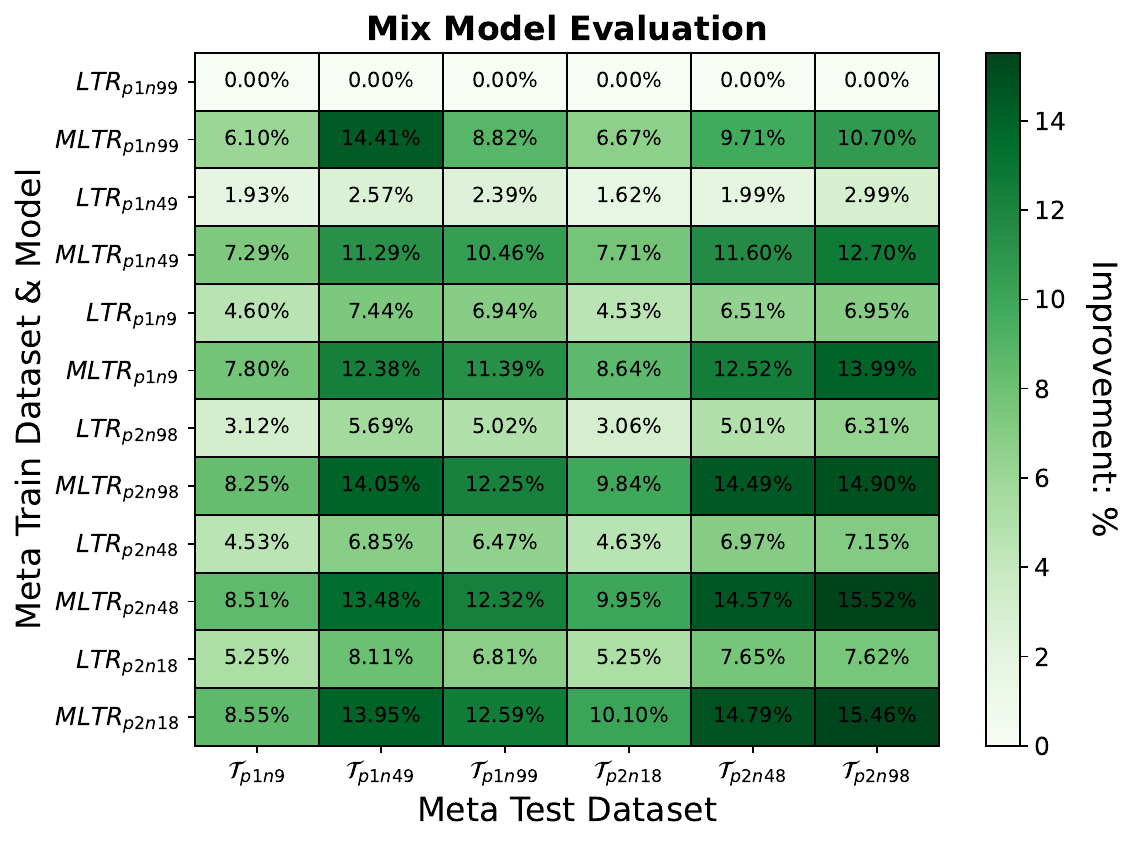}
         \caption{NDCG@5 percentage improvement over $LTR_{p1n99}$}
         \label{fig:casestudy_all}
     \end{subfigure}
     % \hfill
     \begin{subfigure}[b]{0.475\textwidth}
         \centering
         \includegraphics[width=\linewidth]{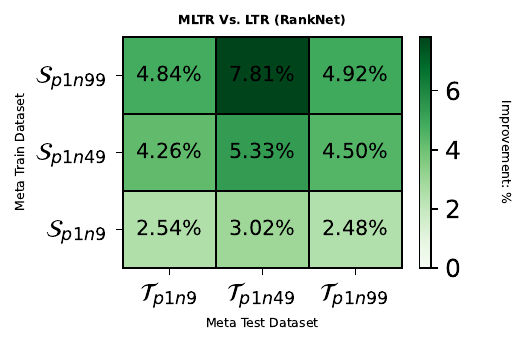}
         \caption{Single positive label comparison}
         \label{fig:casestudy_single}
     \end{subfigure}
     % \hfill
     \begin{subfigure}[b]{0.475\textwidth}
         \centering
         \includegraphics[width=\linewidth]{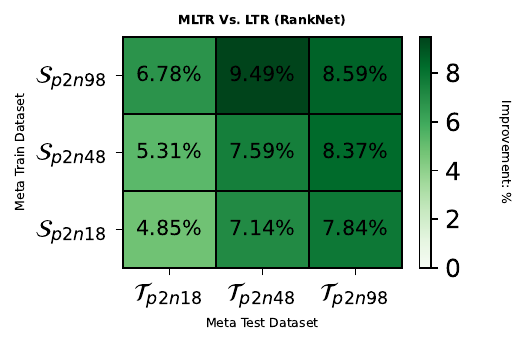}
         \caption{Multiple positive label comparison}
         \label{fig:casestudy_multiple}
     \end{subfigure}
        \caption{Relative improvement experimental results from MLTR and LTR based on RankNet loss in variant sparsely labeled data setting on the Walmart.com dataset}
        \label{fig:casestudy}
\end{figure}

In previous experiments, we simulated sparsely labeled data settings on public datasets. However, the Walmart.com tail query dataset has an even sparser data distribution and more severe data shift issues. Therefore, it is worthwhile to conduct robustness experiments similar to those on public datasets to verify the model's performance on real-world application data. Firstly, we used a configuration similar to the robustness experiments on public datasets and followed the same experimental training and validation process as Section \ref{sec:relative_improvement}. The difference was that we collected a more sparse ratio of positive and negative samples to validate our model's performance in a true application setting.

Fig.~\ref{fig:casestudy_all} shows the relative NDCG@5 gain on the worst-performing model $LTR_{p1n99}$.
For each test data $\mathcal{T}$ (represented by x-axis) corresponding to a column, 
the NDCG@5 metrics are computed for a model ($LTR$ or $MLTR$) trained on $\mathcal{S}$ (represented by y-axis), $LTR_{p \cdot n \cdot}$ and $MLTR_{p\cdot n \cdot}$ denote the model $LTR$ and $MLTR$ train on specific training dataset, separately;
each grid in this column shows the relative NDCG@5 improvement compared against 
$BR_{p1n99}$ on the same test data $\mathcal{T}$. 
The darker the color of each grid, the greater the improvement of the model in this grid relative to $LTR_{p1n99}$.

Firstly, we observed that, similarly to the public dataset experiments, $MLTR$ consistently outperformed $LTR$ for any given $\mathcal{S}$ and $\mathcal{T}$, with a significant improvement of 3.2\% - 10.7\%.
For instance, in the test data $\mathcal{T}_{p1n99}$, $MLTR_{p1n49}$ shows an improvement of 8.07\% over $LTR_{p1n49}$ (as shown in Fig.~\ref{fig:casestudy_all}), with $MLTR$ achieving a 10.46\% improvement compared to a 2.39\% improvement for $LTR$.
Secondly, we noticed that $MLTR$ is considerably more stable and robust in the face of sparse data than $LTR$. This is evident when comparing the performance of all models for a fixed test dataset $\mathcal{T}$ (corresponding to a column), even in the case of more sparse datasets.
As $\mathcal{S}$ gets more sparse, performance degradation is observed for both $MLTR$ and $LTR$ models; 
however, $MLTR$'s NDCG performance decreased much slower compared to $LTR$, 
this observation applies to both scenarios of one positive-labeled item $p1n \cdot$ and two positive-labeled items $p2n \cdot$ per query.
For example, looking at these models' NDCG metrics on $\mathcal{T}_{p1n9}$ (corresponding to column 1) 
as the training data get more sparse from $\mathcal{S}_{p2n18}$ to $\mathcal{S}_{p2n98}$, 
NDCG for $MLTR$ decrease by 2.13\%, from 5.25\%  ($LTR_{p2n18}$) to 3.12\% ($LTR_{p2n98}$), 
while NDCG for $MLTR$ decreases only by 0.3\% from 8.55\% ($MLTR_{p2n18}$) to 8.25\% ($MLTR_{p2n98}$).

We further compared the relative NDCG gain of MLTR over LTR for a combination of $\mathcal{S}$ and $\mathcal{T}$.
Fig.~\ref{fig:casestudy_single} and~\ref{fig:casestudy_multiple} correspond to the cases with 1 positive-labeled item and with 2 positive-labeled item per query respectively. 
Each grid in Fig.~\ref{fig:casestudy_single} and~\ref{fig:casestudy_multiple}
represents the relative NDCG@5 lift of MLTR over LTR when both models are trained on $\mathcal{S}$ (represented by y-axis) and tested on $\mathcal{T}$
(represented by the x-axis). Take the lower right corner grid in Fig. \ref{fig:casestudy_single} as an example, 
it shows a relative 2.48\% improvement of $MLTR_{p1n9}$ over $LTR_{p1n9}$ on the test data $\mathcal{T}_{p1n99}$.
First, looking at models' performance on each test data $\mathcal{T}$ (corresponding to each column),
we see consistent patterns in Fig.~\ref{fig:casestudy_single} and~\ref{fig:casestudy_multiple} 
that the sparser the positive labels in $\mathcal{S}$, the higher the relative NDCG lift of MLTR over LTR. 
Second, fixing the training data $\mathcal{S}$ (corresponding to each row) 
in both Figures~\ref{fig:casestudy_single} and~\ref{fig:casestudy_multiple}, 
we find out that the sparse the test data is,
the more relative improvement of MLTR over LTR overall, 
with an exception of the most extreme case $\mathcal{T}_{p1n99}$.
Third, focusing on the diagonal grids (when $\mathcal{S}$ and $\mathcal{T}$ have the same number of positive/negative labeled items for each query), we see that the sparser the data, the bigger gap between MLTR and LTR in overall except the case of ratio 1:99. 
Overall, the sparser the training data and the test data, the more significant NDCG lift of MLTR over LTR can be observed.
This shows the MLTR model improves the generalization performance over the baseline LTR model on the sparse setting in real-world Walmart dataset.

\section{Conclusion and Future Work}

In this paper, we introduce a novel meta-learning to rank framework that improves the generalization capability of LTR models for search and ranking tasks with sparsely labeled queries. Our proposed model enables quick adaptation to new queries with limited supervision, and produces query-specific rankers with optimal model parameters. Comprehensive experiments demonstrate the versatility and flexibility of our approach, which can be applied to any existing LTR models. Real-world application experiments also further illustrate the effectiveness of our proposed approach.

This work is an initial step toward a promising research direction. First of all, we will explore the application of the proposed framework to neural ranking models and see if the improvement is sustained. We also want to explore strategies for integrating unbiased LTR methods into the training process within the current MLTR framework, further optimize and enhance the model's performance. We will also experiment with larger pre-trained language models as ranking models in our framework. We will study whether meta-learning can provide additional benefit on top of transfer learning. Last but no the least, the proposed framework allows utilization of the meta-learning approach for more real-world applications, such as head-to-tail transfer learning, as well as the multi-task learning problem, etc. 

\begin{acks}
We would like to express our deepest gratitude to the team at Walmart Global Tech for providing us with the opportunity to undertake this research project. Their guidance, expertise, and resources were invaluable in helping us to complete this work. 
\end{acks}

\bibliographystyle{ACM-Reference-Format}
\bibliography{reference}

\end{document}